\documentclass[11pt]{article}

\usepackage[utf8]{inputenc}
\usepackage[T1]{fontenc}
\usepackage[a4paper, nohead]{geometry}
\usepackage{amsmath,amssymb} 
\usepackage{graphicx}
\usepackage[affil-it]{authblk}
\usepackage{enumitem}
\usepackage{empheq}
\usepackage{pdfpages}
\usepackage{cite}
\usepackage{adjustbox}
\usepackage[british]{babel}

\usepackage[colorlinks,linkcolor=blue,citecolor=blue]{hyperref}

\numberwithin{equation}{section}
\numberwithin{footnote}{section}

\newtheorem{defn}{Definition}

\newcounter{example}[section]

\newenvironment{case}{%
	\let\olditem\item% 
	\renewcommand\item[1][]{\olditem \textbf{##1} \\}%
	\begin{enumerate}[label=\textbf{Case \arabic*:},itemindent=*,leftmargin=2em]}{\end{enumerate}%
}

\usepackage[titletoc]{appendix}

\newcommand{\R}{\mathbb{R}}
\newcommand{\keyword}[1]{\emph{#1}}
\newcommand{\Eqref}[1]{Eq.~\eqref{#1}}
\newcommand{\Eqsref}[1]{Eqs.~\eqref{#1}}
\newcommand{\Sectionref}[1]{Section~\ref{#1}}

\newcommand{\Figref}[1]{Fig.~\ref{#1}}

\newcommand{\kstar}{\overset{\star}{k}}

\title{Numerical construction of initial data sets  of binary black hole type using a parabolic-hyperbolic formulation of the vacuum  constraint equations}

\author[1]{F. Beyer\footnote{Email: fbeyer@maths.otago.ac.nz.} }
\author[2]{L. Escobar\footnote{Email: leon.escobar@correounivalle.edu.co.} }
\author[1]{J. Frauendiener \footnote{Email: joergf@maths.otago.ac.nz.} }
\author[1]{J. Ritchie \footnote{Email: jritchie@maths.otago.ac.nz}}
\affil[1]{Department of Mathematics and Statistics, University of Otago,  New Zealand.}
\affil[2]{Departmento  de  Matematicas, Universidad del Valle,  Colombia.}
\date{\today}

\begin{document}

\maketitle

\begin{abstract}
In this paper we investigate the parabolic-hyperbolic formulation of the vacuum constraint equations introduced by R{\'a}cz with a view to constructing multiple black hole initial data sets without spin. In order to respect the natural properties of this configuration, we foliate the spatial domain with $2$-spheres. It is then a consequence of these equations that they must be solved as an initial value problem evolving outwards towards spacelike infinity. Choosing the free data and the ``strong field boundary conditions'' for these equations in a way which mimics asymptotically flat and asymptotically spherical binary black hole initial data sets, our focus in this paper is on the analysis of the asymptotics of the solutions. In agreement with our earlier results, our combination of analytical and numerical tools reveals that these solutions are in general not asymptotically flat, but have a cone geometry instead. In order to remedy this and approximate asymptotically Euclidean data sets, we then propose and test an iterative numerical scheme.
\end{abstract}

%%%%%%%%%%%%%%%%%%%%%%%%%%%%%%%%%%%%%%%%%%%%%%%%%%%%%%%%%%%%%%%%%%%%%%%%%%%%%%%%%%%%%%%%%%%%%%%%%%
%-------------------------------------------------------------------------------------------------

\section{Introduction}\label{Sec:Introduction}

 Thanks to the ground-breaking work by Choquet-Bruhat and Geroch \cite{FouresBruhat:1952ji,ChoquetBruhat:1969cl} we know that each solution of the Einstein vacuum constraint equations (an \emph{initial data set}, see below)  determines a unique solution of the full vacuum Einstein's equations -- the so-called \emph{maximal globally hyperbolic development} -- in which the initial data set arises as the induced geometry of some spacelike hypersurface. 
A major concern of modern general relativity research is therefore how to construct solutions of these constraints. In all of what follows, a triple $(\Sigma,\gamma_{a b},K_{ab})$ of a $3$-dimensional differentiable manifold $\Sigma$, a Riemannian metric $\gamma_{a b}$ and a smooth symmetric tensor field $K_{ab}$ on $\Sigma$ is called an \emph{initial data set} if it satisfies the \emph{vacuum constraint equations}\footnote{All physical units in this paper are geometric ($G=c=1$), i.e., distances, time and mass are measured with the same physical unit.}  
\begin{equation}\label{eqs:constraints_equations}
	R-K_{ab}K^{ab}+K^{2}=0,\;\; \nabla_{a}{K^{a}}_{c}-\nabla_{c}K=0
	\end{equation} 
everywhere on $\Sigma$. Here, $\nabla_a$ is the Levi-Civita covariant derivative determined by $\gamma_{a b}$, $R$ the Ricci scalar
and $K = {K^{a}}_{a}$. For this whole paper we agree that spatial abstract indices $a,b,\ldots$ are raised and lowered with the metric $\gamma_{a b}$.

In general, the physical scenario of interest determines which particular properties one is looking for in solving  \Eqref{eqs:constraints_equations}. In this paper here for example we are interested in isolated systems of vacuum black holes.
 However, the constraints 
form an under-determined system and, to the best of our knowledge, it is not currently known if there exists a geometrically or physically preferable way to decide what part of the data should be specified and what part should be solved for. One of the most widely used approaches, both analytically and numerically, is that by Lichnerowicz and York (see \cite{bartnik2004, Baumgarte:2010vs} and references therein) which involves specifying the conformally covariant ingredients of the data and solving a  non-linear elliptic boundary value problem equivalent to the constraints. There are several well-established numerical techniques \cite{Baumgarte:2010vs} for solving this boundary value problem.

While this approach by Lichnerowicz and York is very successful both analytically and numerically, it has also been criticised on several grounds in the literature. There are  analytical problems when solutions are sought with highly non-constant mean curvature (see \cite{dilts2017,anderson2018a} for an overview and references). There are physical problems related to ``spurious radiation'' (see e.g.\ \cite{chu2014}, but also \cite{lovelace2009}). While  discussions of such issues are  ongoing  in the literature, they have motivated several authors \cite{Bishop:1998cb,Matzner:1998hv,Bishop:1998cb,Moreno:2002dm,Bishop:2004gb} to strive for alternative approaches for solving the constraints. The particular approach of interest for this paper here was introduced by R\'acz in a series of papers \cite{Racz:2014kk,Racz:2014dx, Racz:2015bu,Racz:2015gb} following earlier work by Bartnik \cite{bartnik1993}. As we discuss below in more detail, the constraint equations are turned into either a hyperbolic-algebraic system or  a parabolic-hyperbolic system depending on which components of the data are regarded as free. These systems are to be solved as an \emph{initial value problem}. 

First steps in investigating whether this approach has any advantages over more established methods  have been carried out in \cite{racz2018} for the constraints of the Maxwell equations and in \cite{Nakonieczna:2017vk,doulis2019} for the Einstein constraint equations. 
One obvious issue with solving the constraints as an initial value problem is the lack of control of the properties of the solutions either in the strong field regime or at spacelike infinity -- two regimes which crucially characterise the physical properties of the resulting data set. Winicour considered the hyperbolic-algebraic formulation to obtain linearised perturbations of the Minkowski spacetime \cite{winicour2017}. Somewhat disappointingly, he found that there are no suitable Cauchy hypersurfaces in the Minkowski spacetime on which the linearised algebraic-hyperbolic formulation of the constraints are well posed.  This paper here now is a continuation of the investigations, which we started in \cite{Beyer:2017tu} for the hyperbolic-algebraic formulation, of the asymptotics at spacelike infinity for the parabolic-hyperbolic one. In order to understand which questions are of interest to us, let us note that
any initial data set should have some reasonable asymptotic behaviour at spacelike infinity if it was to represent black holes. Without this fundamental notions associated with black holes, like event horizons and masses, and with gravitational radiation may not even be defined (see, for example, \cite{Szabados:2009ig}). The most prominent requirement, which we shall consider in this paper, is  \emph{asymptotic flatness}.
\begin{defn}
    \label{AsympFlatDef}    
According to \cite{Dain:2001cd} we say that an initial data set $(\Sigma,\gamma_{ab},K_{ab})$ is \emph{asymptotically flat} if $\Sigma$ is diffeomorphic to $\R^3$ (possibly minus a ball of finite size) and  if there exist coordinates $\{\tilde{x}^{i}\}$ on $\Sigma$ such that the components of $\gamma_{ab}$ and $K_{ab}$ with respect to these coordinates satisfy\footnote{\label{ftn:OSymbol1} We use the $O$ symbol rather informally in the usual sense $f = O(g) \iff |f| \le C |g| $ for some constant $C>0$ in the relevant limit. In the case of tensors, we require that each coordinate component of the tensor satisfies such an estimate with respect to some natural coordinate system (like Cartesian coordinates in Definition~\ref{AsympFlatDef}). In this paper, we avoid the technicalities of choosing appropriate norms for the definition of the $O$-symbol. In fact, in order to make the above notions of asymptotic flatness precise \emph{and} physically meaningful, the $O$-symbol must be defined with respect to a norm which does not only control the decay of the fields themselves, but also of an appropriate number of derivatives. If this is the case, asymptotic flatness in the sense above can be shown to imply that the curvature tensor  decays at infinity. The interested reader can find  details in the references \cite{Szabados:2009ig,Dain:2001cd}.}
\begin{align}  
        \gamma_{ij}&=\left( 1 + \frac{2M}{R} \right)\delta_{ij}+{O}(R^{-2}),  \label{FlatLimit2}\\
         K_{ij}&= O( R^{-2} ) \label{FlatLimit3},             
      \end{align} 
      in the limit 
        \begin{equation}
        R=\left(\sum_{a=1}^{3} (\tilde{x}^{i})^2 \right)^{1/2} \rightarrow \infty,
        \label{FlatLimit}
      \end{equation}
Here $\delta_{ab}$ denotes the Euclidean metric and $M\ge 0$ is the \emph{ADM mass}.
\end{defn}
If  \Eqref{FlatLimit3} fails, but the other conditions are satisfied, the initial data is called  
\emph{asymptotically Euclidean}. Regarding other notions of asymptotic flatness, which are useful in different contexts, see \cite{bieri2009extensions,klainerman1993global}.

In \cite{Beyer:2017tu} the asymptotics of a class of initial data sets constructed by means of  R\'acz's {algebraic-hyperbolic} formulation of the constraints was investigated for perturbations of a \emph{single} Schwarzschild black hole. It was found that, in general, these initial data sets violate asymptotic flatness. In this paper now, we continue  these explorations by investigating the {parabolic-hyperbolic} formulation in a \emph{multiple} black hole setting. One of the key ingredients of R\'acz's approach is that space is foliated in a $2+1$-fashion. While the topology of the $2$-dimensional leaves of the foliation is largely free, a particularly natural choice to capture an isolated black hole configuration, which we adopt in all of what follows, are distorted $2$-spheres centred in the strong field regime around the black holes (in contrast to the choice of $2$-planes in \cite{Nakonieczna:2017vk,doulis2019}). Because of this choice, we can also reuse the numerical pseudo-spectral methods developed in \cite{Beyer:2017jw,Beyer:2014bu,Beyer:2016fc,Beyer:2009vw} which were also employed in \cite{Beyer:2017tu}. As we discuss in detail, it is then a \emph{consequence} of the {parabolic-hyperbolic} equations by R\'acz that they must be solved as an initial value problem starting from a boundary in the strong field regime evolving outwards towards spacelike infinity.
Similar to the findings in \cite{Beyer:2017tu}  the resulting data sets turn out to violate asymptotically flatness; in fact, they are in general not even asymptotically Euclidean. In order to remedy this, we then introduce an iterative procedure which we demonstrate to allow us to approximate asymptotically Euclidean initial data sets. This paper exclusively focusses on the asymptotics at spacelike infinity. We have not performed any searches for apparent horizons or performed any other investigations of the properties of our data sets in the strong field regime.

Our paper is organised as follows. In 
\Sectionref{Sec:ThevacuumConstraintsasaParabolic-HyperbolicSystem} we briefly
summarise R\'acz's para\-bolic-hyperbolic formulation of the constraint equations while \Sectionref{sec:KSID} introduces the Kerr-Schild formalism which turns out to be useful for us later. In
\Sectionref{Sec:Binary_black_hole_model} we consider some explicit solutions of the parabolic-hyperbolic system and analyse a family of exact spherically symmetric solutions in order to set the scene for our analysis in the
following sections. We present a brief discussion of Bishop's superposition
method \cite{Bishop:1998cb} and then adapt it to the parabolic-hyperbolic formulation of the constraints. We discuss the numerical methods involved in solving the system of equations. In 
\Sectionref{Sec:Asymptotics_of_binary_black_hole_data} we
analyse the asymptotic behaviour of the numerical solutions and introduce an iterative procedure for constructing asymptotically Euclidean initial data sets. Finally, in 
\Sectionref{Sec:Conclusion}, we summarise our main findings.

%%%%%%%%%%%%%%%%%%%%%%%%%%%%%%%%%%%%%%%%%%%%%%%%%%%%%%%%%%%%%%%%%%%%%%%%%%%%%%%%%%%%%%%%%%%%%%%%%%
%-------------------------------------------------------------------------------------------------
\section{Preliminaries}
\subsection{The parabolic-hyperbolic formulation of the constraints}
\label{Sec:ThevacuumConstraintsasaParabolic-HyperbolicSystem}

In this section we briefly summarise R\'acz's parabolic-hyperbolic formulation of the vacuum constraint equations. Further details can be found in \cite{Racz:2014kk,Racz:2014dx,Racz:2015gb,
Racz:2015bu}. Since our conventions and notation partly deviate from those in these earlier references we provide a table in Appendix \ref{A_NotationConvention} comparing the different conventions.

Consider any {data set} $(\Sigma,\gamma_{ab},K_{ab})$ where $(\Sigma, \gamma_{ab})$ is  a $3$-dimensional Riemannian manifold and $K_{ab}$ is a smooth symmetric tensor field on $\Sigma$. The Levi-Civita covariant derivative associated with $\gamma_{ab}$ is denoted by $\nabla_a$. Suppose in addition that there is a smooth function $\rho:\Sigma\rightarrow\R$ whose level sets $S_\rho$ are smooth $2$-dimensional surfaces in $\Sigma$ and that the collection $S$ of all these surfaces is a foliation of $\Sigma$. The  quantity $t^a \nabla_a\rho$ vanishes for all vectors $t^a$ tangent to the surfaces $S_\rho$ and 
\begin{equation}
  \label{eq:defNa}
  N_a=A \nabla_a\rho
\end{equation}
is the unit co-normal where $A$ is a strictly positive smooth function called the \emph{lapse} (function).

Now we decompose the data set $(\Sigma,\gamma_{ab},K_{ab})$ with respect to the foliation defined by the function $\rho$ in full analogy to the standard $3+1$-decomposition of spacetimes, see for example \cite{Baumgarte:2010vs}.

The first and second fundamental forms of the surfaces $S_\rho$ are 
\begin{equation}
\label{eq:defhab}
h_{ab}=\gamma_{ab}-N_{a}N_{b},
\end{equation}
and\footnote{Observe our sign convention for second fundamental forms. As outlined in the table in Appendix \ref{A_NotationConvention} this choice of sign is different in most of the earlier literature.}
\begin{equation}
  \label{eq:defkab}
    k_{ab}=-\frac{1}{2}\mathcal{L}_{N}h_{ab},
\end{equation}
respectively. 
The {orthogonal projector} onto the surfaces is therefore
\begin{equation*}
{h^a}_{b}={\delta^a}_{b}-N^{a}N_{b}.
\end{equation*}
The covariant derivative associated with $h_{ab}$ is referred to as $D_a$.

We say that a tensor field on $\Sigma$ is \textit{intrinsic (to the surfaces $S_{\rho}$)} if any contraction with $N_{a}$ or $N^{a}$ vanishes. In particular,  $h_{ab}$ and $k_{ab}$ are both intrinsic. Contracting all indices of a tensor field with ${h^a}_{b}$ always yields an intrinsic tensor field. In fact, any tensor field can be decomposed uniquely into intrinsic and normal parts as follows
\begin{equation}\label{eq:splittingofsecondfundamentalform}
  K_{ab}=\kappa N_{a}N_{b} + N_{a}p_{b} + N_{b}p_{a} + q_{ab},
\end{equation}
where we require that $p_a$ and $q_{ab}$ are intrinsic. The symmetric intrinsic tensor field $q_{ab}$  can then be decomposed into its trace and trace-free part (with respect to $h_{ab}$)
\begin{equation}
\label{eq:splitqab}
q_{ab}=Q_{ab}+\frac{1}{2}q h_{ab},\quad Q_{ab} h^{ab}=0.
\end{equation}

Next we pick an arbitrary smooth vector field $\rho^{a}$ normalised by the condition
\begin{equation}
 \rho^a  \nabla_a  \rho = 1.
\end{equation}
Due to \Eqref{eq:defNa}, this means that there must exist an intrinsic vector field $B^a$, the \emph{shift} (vector), such that 
\begin{equation}
  \label{eq:defB}
 \rho^a = A N^a + B^a.
\end{equation}
Given $\rho^a$, we can write $k_{ab}$ in \Eqref{eq:defkab} as
\begin{equation}
\label{eq:defkStar}
k_{ab}=-A^{-1}\left(\frac{1}{2}\mathcal{L}_{\rho}h_{ab}-
D_{\left( a\right.}B_{\left. b\right)} \right)=: A^{-1} \overset{\star}{k}_{ab}.
\end{equation}
We set
\begin{equation}
  \label{eq:defkStar2}
  \kstar=h^{ab}\overset{\star}{k}_{ab}.
\end{equation}
As found in \cite{Racz:2015gb} and as we will discuss below, the crucial property of the quantity $\overset{\star}{k}_{ab}$ is that it only depends on $\rho^a$, $h_{ab}$ and $B_a$ (the ``free data'' for the constraint equations) and not on $A$ (one of the ``unknowns'').
Finally we use \Eqref{eq:defNa} to write the intrinsic acceleration as
\begin{equation}
  \label{eq:defvb}
  v_{a}:=N^{b}\nabla_{b}N_{a}=-A^{-1}D_{a}A.
\end{equation}
    
According to \cite{Racz:2015gb} the constraint equations \eqref{eqs:constraints_equations} now take the following form by decomposing the momentum constraint covector into its normal and intrinsic parts and performing appropriate manipulations of the Hamiltonian constraint\footnote{These equations agree with \cite{Racz:2014kk}, when the matter terms have been set to zero. }:
 \begin{align}
\overset{\star}{k}\nabla_{\rho}A+A^2 D^{a}D_a A-\overset{\star}{k}B^a D_{a}A=&
\frac 12 A^3 E
+\frac 12A F,
\label{ParabolicEquation}
\\
\nabla_{\rho}q-B^a D_{a}q-AD_{a}p^{a}-2p^{a}D_{a}A=&\overset{\star}{k} {}^{ab}Q_{ab}+\frac{1}{2}q\overset{\star}{k} -\overset{\star}{k}\kappa,
\label{FinalSystemDiffNorm}
\\
{h^b}_c\nabla_{\rho}p_{b}-B^a D_{a}p_{c}-\frac{1}{2}AD_{c}q=&AD_{c}\kappa-AD_{a} {Q^a}_{c} - {Q^a}_{c}D_{a}A-\frac{1}{2}qD_{c}A 
\label{FinalSystemDiffMom}
\\
&+\overset{\star}{k}p_c  +p^{b}\overset{\star}{k}_{bc}+\kappa  D_{c}A,
  \nonumber
\end{align}
where $\nabla_\rho=\rho^a\nabla_a$, and,
\begin{equation*}
E={}^{(2)}R+2\kappa q-2p^{a}p_{a}-Q_{ab}Q^{ab}+\frac{1}{2}q^2,
\quad
F=2(\nabla_{\rho}\overset{\star}{k}-B^a D_{a}\overset{\star}{k})-
\overset{\star}{k}_{ab}{\overset{\star}{k}} {}^{ab}-\overset{\star}{k}{}^2.
\end{equation*}
The Ricci scalar associated with  $h_{ab}$ is called ${}^{(2)}R$.
Notice that $\nabla$ can be eliminated in favour of Lie-derivatives. In \Eqsref{ParabolicEquation} and \eqref{FinalSystemDiffNorm} we can write $\mathcal L_{\rho} A$ instead of $\nabla_\rho A$ and $\mathcal L_{\rho} q$ instead of $\nabla_\rho q$, and
 \Eqref{FinalSystemDiffMom} can be written as
\begin{equation}
  \begin{split}
\mathcal L_{\rho} p_b-B^a D_{a}p_{c}-\frac{1}{2}AD_{c}q=\,&p_a D_b B^a
-AD_{a} {Q^a}_{c}+\kappa  D_c A -{Q^a}_{c}D_{a}A-\frac{1}{2}qD_{c}A
\\
&+\overset{\star}{k}p_c + AD_{c}\kappa.
\end{split}
\end{equation}

These equations suggest to group the various fields introduced above as follows:
\begin{description}
\item[Free data] The fields $B_{a}$, $Q_{ab}$,  $h_{ab}$ and $\kappa$ are considered as {freely specifiable} in  \eqref{ParabolicEquation}--\eqref{FinalSystemDiffMom} everywhere
  on $\Sigma$. Notice from the above that $\kstar$, $D_a$,
  ${}^{(2)}R$, $Q_{ab}$ and $F$ (and all index versions of these
  intrinsic fields) can be calculated from the free data everywhere on
  $\Sigma$.
\item[Unknowns] The fields $A$, $q$ and $p_{a}$ are considered as unknowns which one attempts to determine as solutions of \eqref{ParabolicEquation}--\eqref{FinalSystemDiffMom}. Notice that all coefficients in these equations can be calculated from the free data everywhere on $\Sigma$.
\end{description}
    According to \cite{Racz:2014kk}, it can be shown that given any smooth \emph{initial data}\footnote{The initial datum for $A$ is assumed to be strictly positive.} for $A$, $q$ and $p_a$ on any $\rho=\rho_0$-leaf of the $2+1$-decomposition of $\Sigma $ in addition to smooth \emph{free data} everywhere $\Sigma$, the \emph{initial value problem} of \Eqsref{ParabolicEquation} -- \eqref{FinalSystemDiffMom} in the \emph{increasing} $\rho$-direction is well-posed, i.e., the equations have a unique smooth solution $A$, $q$ and $p_{a}$ at least in a neighbourhood of the initial leaf, provided the \keyword{parabolicity condition}\footnote{The table in Appendix \ref{A_NotationConvention} explains the sign discrepancy with \cite{Racz:2015gb}.} holds everywhere on $\Sigma$:
\begin{equation}
  \label{eq:parabolcond}
  \kstar<0.
\end{equation}
We remark that if $\kstar$ is positive instead, then the {initial value problem} in the \emph{decreasing} $\rho$-direction is well-posed instead.
In either case, \Eqsref{ParabolicEquation} -- \eqref{FinalSystemDiffMom} is a quasilinear parabolic-hyperbolic system. It is important to notice that $\kstar$ is fully determined by the free data. \Eqref{eq:parabolcond} can therefore be checked \emph{before} one attempts to solve \Eqsref{ParabolicEquation} -- \eqref{FinalSystemDiffMom}. We remark that in the context of black hole data sets, the initial data for \eqref{ParabolicEquation}--\eqref{FinalSystemDiffMom} imposed on the initial $\rho=\rho_0$-leaf of the $2+1$-decomposition is often referred to as \emph{boundary conditions in the strong field regime} in the literature.

As in \cite{Beyer:2017tu} we restrict to the case $\Sigma=\R^3\backslash \overline B$ where $B$ is some finite ball in $\R^3$ in all of what follows. Moreover, we assume  that the level sets of $\rho$ are diffeomorphic to $2$-spheres. Following  \cite{Penrose:1984tf,Beyer:2015bv,Beyer:2014bu,Beyer:2016fc,Beyer:2017jw,Beyer:2017tu}, all \emph{intrinsic} tensor fields can therefore be written in terms of quantities with well-defined \keyword{spin-weights} (see \Sectionref{Sec:SWSHstuff} in the appendix for a quick summary). We can also express the intrinsic covariant derivative operator $D_a$ (defined with respect to the intrinsic metric $h_{ab}$)  in terms of the covariant operator $\hat D_a$ defined with respect to the round unit-sphere metric $\Omega_{ab}$; recall that $D_a-\hat D_a$ can be expressed by some smooth intrinsic tensor field. Using \Sectionref{Sec:SWSHstuff}, we can then express the covariant derivative operator $\hat D_a$ in terms of the $\eth$- and $\eth'$-operators \cite{Penrose:1984tf}. Once all of this has been completed for all terms in \Eqsref{ParabolicEquation} -- \eqref{FinalSystemDiffMom}, each of these equation and each term ends up with a consistent well-defined spin-weight. Most importantly, however, all terms are explicitly regular: Standard polar coordinate issues at the poles of the $2$-sphere disappear when all quantities are  expanded in terms of \keyword{spin-weighted spherical harmonics} and \Eqsref{eq:eths} and \eqref{eq:eths2} are used to calculate the intrinsic derivatives.
From the numerical point of view this gives rise to a (pseudo-)spectral scheme. Further details related to our implementation can be found in \cite{Beyer:2017tu}.

\subsection{Data sets of Kerr-Schild form}
\label{sec:KSID}
In this subsection we introduce general Kerr-Schild-like data sets. These will play an important role in the remainder of this paper. We stress that we do not yet impose the constraint equations in this subsection. Sometimes we call  data sets \keyword{unphysical} or \keyword{preliminary} if the constraints are not imposed. 

Let us start our discussion with a $4$-dimensional Lorentzian manifold\footnote{Note that we use Greek indices for denoting  spacetime abstract indices. Spacetime indices  are consistently raised and lowered with $g_{\alpha\beta}$.} $(M,g_{\alpha\beta})$ where the metric $g_{\alpha\beta}$ is of Kerr-Schild form
\begin{equation}
g_{\alpha\beta}=\eta_{\alpha\beta}-Vl_{\alpha}l_{\beta}.
\label{GeneralKerr}
\end{equation}
Here $\eta_{\alpha\beta}$ is the Minkowski metric, $l^\alpha$ is a null vector field with respect to $g_{\alpha\beta}$ and $V$ is a smooth spacetime function. Notice that
$l_{\alpha}=g_{\alpha\beta}l^\beta=\eta_{\alpha\beta}l^\beta$
and that $l^\alpha$ is therefore automatically also null with respect to $\eta_{\alpha\beta}$. We also assume coordinates $(t,x,y,z)$ such that $\eta_{\alpha\beta}$ takes the form $\mathrm{diag}(-1,1,1,1)$ and normalise $l^\alpha$ such that its $t$-component is $1$. 

Given this it turns out that the $t=\mathrm{const}$-surfaces $\Sigma_t$ are spacelike if $V<1$, as we shall always assume. 
The future directed  (with respect to the $\partial_t$) unit normal is\footnote{Coordinate one-forms, e.g.\ $dt$, are either written in index-free notation, or, in abstract index notation $dt_\alpha$. Notice that $dt_\alpha=\overline{\nabla}_\alpha t$ where $\overline\nabla_\alpha$ is the covariant derivative associated with $g_{\alpha\beta}$. Similarly we write either $\partial_t$ or $\partial^\alpha_t$ for coordinate vector fields.} $n_\mu= -dt_\mu/\sqrt{1-V}$. The spacetime lapse function is therefore $\alpha=1/\sqrt{1-V}$, the induced metric 
$\gamma_{\mu\nu}=g_{\mu\nu}+dt_\mu dt_\nu/(1-V)$,
the shift is $\beta^\mu=\partial_t^\mu-\alpha n^\mu$ and the second fundamental form is $K_{\mu\nu}=-\frac{1}{2} \mathcal L_n \gamma_{\mu\nu}$. For any fixed $t\in\R$, consider now the embedding $\Phi: \Sigma\rightarrow\Sigma_t$, $(x,y,z)\mapsto (t,x,y,z)$ where $\Sigma$ is a smooth $3$-dimensional manifold diffeomorphic to $\Sigma_t$. Adopting the index notation introduced in \Sectionref{Sec:ThevacuumConstraintsasaParabolic-HyperbolicSystem} (in particular, all spatial index operations are  performed with $\gamma_{ab}$ as before), we denote the pull-backs of the above quantities from $\Sigma_t$ to $\Sigma$ as $\gamma_{ab}$, $K_{ab}$, $l_a$, $\beta_a$ and $\alpha$, respectively. We find straightforwardly
  \begin{align}
    \label{eq:KS1}
    \gamma_{ab}&=\delta_{ab}-V l_a l_b,\\
    \alpha&=\frac 1{\sqrt{1-V}},\quad
    \beta_a=V l_a,\\
    \label{eq:KS3}
    K_{ab}&=\frac{1}{2\alpha}\left(\nabla_a \beta_b+\nabla_b\beta_a-\mathcal L_t \gamma_{ab}\right)
    =\frac{1}{2\alpha}\left(\nabla_a (V l_b)+\nabla_b(V l_a)-\mathcal L_t \gamma_{ab}\right),
  \end{align}
  where $\delta_{ab}$ is the Euclidean metric which in Cartesian coordinates $(x,y,z)$ takes the form $\text{diag}(1,1,1)$. The field $l_a$ has magnitude $1$ with respect to $\delta_{ab}$, i.e., 
if we set\footnote{Observe carefully that $\delta^{ab}$ and $(\delta^{-1})^{ab}$ are different fields: The first one is defined by raising both indices of $\delta_{ab}$ with $\gamma^{ab}$, while the second one is the uniquely determined inverse of $\delta_{ab}$.}
\begin{equation}
  \tilde{l}^{a}=(\delta^{-1})^{ab}l_{b},
\end{equation}
then 
\begin{equation}
  \label{eq:lnorm}
  \tilde{l}^{a}l_a=(\delta^{-1})^{ab}l_a l_b=1. 
\end{equation}
In particular, it follows that
\[\gamma^{ab}=(\delta^{-1})^{ab}+\frac{V}{1-V}\tilde{l}^{a}\tilde{l}^{b}\]
and
\begin{equation}
  \label{eq:normalla}
  l^a=\frac 1{1-V}\tilde{l}^{a},\quad l^al_a=\frac 1{1-V}.
\end{equation}

\newcommand{\normall}{f}
Now pick a  smooth function $\rho$ on $\Sigma$ with the properties discussed in  \Sectionref{Sec:ThevacuumConstraintsasaParabolic-HyperbolicSystem} giving rise to a foliation $S$ in terms of  level sets $S_\rho$. We restrict to the case where $l_a$ is normal to $S_\rho$, i.e.,
\begin{equation}
  \label{eq:specla}
  l_a=\pm\normall\nabla_a\rho,
\end{equation}
with
\begin{equation}
  \label{eq:lanorm}
  \normall=\frac 1{\sqrt{(\delta^{-1})^{ab}\nabla_a\rho\nabla_b\rho}}.
\end{equation}
From \Eqsref{eq:defNa}, \eqref{eq:specla} and \eqref{eq:normalla} we find that
\begin{equation}
  N_a=\sqrt{1-V}\, l_a,
\end{equation}
which means that the lapse defined in \Eqref{eq:defNa} is
\begin{equation}
  A=\normall \sqrt{1-V}.
\end{equation}
Given \Eqsref{eq:defhab} and \eqref{eq:KS1}, it follows that
\begin{equation}
  \label{eq:KShab}
  h_{ab}=\delta_{ab}-l_a l_b.
\end{equation}
Plugging all this into \Eqref{eq:KS3} yields
\begin{equation}
\label{eq:KSKab}
  K_{ab}
  =\frac{2-V}{4(1-V)}\left(\nabla_a V N_b+\nabla_b V N_a\right)
    +\frac{V}{2}\left(\nabla_a N_b+\nabla_b N_a\right)
    -\frac{\sqrt{1-V}}{2}\dot\gamma_{ab},
\end{equation}
where 
\begin{equation}
  \label{eq:gendotgamma}
  \dot\gamma_{ab}=\mathcal L_t \gamma_{ab}
\end{equation}
and $t^\mu=\partial_t^\mu$ has the coordinate representation $(1,0,0,0)$. 
From \Eqref{eq:splittingofsecondfundamentalform} we get
\begin{align}
  \label{eq:KSkappa}
  \kappa&=\frac{2-V}{2(1-V)^{3/2}} \tilde l^a\nabla_a V 
    -\frac{\sqrt{1-V}}{2}\dot\gamma_{ab}N^aN^b,\\
  \label{eq:KSpa}
  p_a&=\frac{2-V}{4(1-V)} D_a V 
    +\frac{V}{2}v_a -\frac{\sqrt{1-V}}{2}\dot\gamma_{cb}{h^{c}}_{a}N^b,\\
  \label{eq:KSqab}
  q_{ab}&=-V {k}_{ab}-\frac{\sqrt{1-V}}{2}\dot\gamma_{cd}{h^{c}}_{a}{h^{d}}_{b},
\end{align}
where $v_a$ can be calculated from \Eqref{eq:defvb} and $k_{ab}$ from \Eqref{eq:defkStar}. 
The quantities $q$ and $Q_{ab}$ are given by
\Eqref{eq:splitqab}.

Once we have picked intrinsic coordinate systems $(y^1,y^2)$ of the $\rho=\mathrm{const}$-surfaces and thereby an ``adapted'' coordinate system $(\rho,y^1,y^2)$ of $\Sigma$, the vector field $\rho^a=\partial_\rho^a$ is determined. The shift $B_a$ in \Eqref{eq:defB} is then given as
\[
  B_a=\rho^bh_{ab}
\]
and $k_{ab}$, $\overset{\star}{k}_{ab}$ and $\kstar$ can be calculated from \Eqsref{eq:defkStar} and \eqref{eq:defkStar2}.

A particularly important example is the Kerr-Schild Schwarzschild initial data set with mass $M\in\R$. It can be written  in Kerr-Schild form as (see \cite{Alcubierre:Book,Misner:1973vb})\footnote{In general one could have $l_\mu=-dt_{\mu}\pm dr_\mu$ which distinguishes \emph{ingoing} Kerr-Schild coordinates from \emph{outgoing} ones. We use the positive sign exclusively here.}
\begin{equation}
\label{eq:SSKS}
V= -\frac{2M}{r},\quad l_{\mu}
=-dt_{\mu}+dr_\mu,
\end{equation}
where $r=\sqrt{x^2+y^2+z^2}$. Using the formalism discussed above, we find that
\[l_a=\nabla_a r,\quad\dot\gamma_{ab}=0.\]
It is therefore consistent with the assumption \Eqref{eq:specla} when we identify the function $\rho$ with the coordinate $r$, whose level sets are $2$-spheres.
Straightforward calculations yield
\begin{gather}\label{KerrSchildFreeData}
B_{a}=0,\;\;\, Q_{ab}=0, \;\;\, \kappa=\frac{2M(M+r)}{r^3\, \lambda(r)^3 },\;\;\, h_{ab} = r^2 \left( d\theta_{a} d\theta_{b} + \sin^{2}\theta\, d\phi_{a} d\phi_{b} \right),\\
\label{KerrSchildSolutions}
\kstar = -\dfrac{2}{r},\quad {p}_{a}=0,\;\;\, {q}= -\frac{4 M}{r^2 \, \lambda(r)},\;\;\, {A} = \lambda(r) , 
\end{gather}
where $(\theta,\phi)$ are standard polar coordinates on each $r=\mathrm{const}$-sphere and where we defined 
\begin{align}
\lambda(r) = \sqrt{1+\frac{2M}{r}}.
\label{Eq:Def_Lambda}
\end{align} 
Observe that the parabolicity condition \Eqref{eq:parabolcond} is  satisfied for all $r>0$.

%%%%%%%%%%%%%%%%%%%%%%%%%%%%%%%%%%%%%%%%%%%%%%%%%%%%%%%%%%%%%%%%%%%%%%%%%%%%%%%%%%%%%%%%%%%%%%%%%%
%-------------------------------------------------------------------------------------------------
\section{Multiple black hole initial data sets}
\label{Sec:Binary_black_hole_model}

\subsection{Outline of our approach}

In order to construct multiple black hole initial data sets using the formalism introduced in \Sectionref{Sec:ThevacuumConstraintsasaParabolic-HyperbolicSystem}
we proceed now in two steps. The first step is to produce, without imposing the constraints yet, data sets which can somehow be interpreted as multiple black holes. As discussed in \Sectionref{SubSec:ConstructingAuxiliaryData} we approach this in a largely ad hoc and certainly not unique way. Because the constraints are not imposed yet, such data sets are referred to as (preliminary) data sets. Only in the second step, see \Sectionref{Sec:Numerical_implementation_and_tests}, such a preliminary data set is used to obtain the \keyword{free data} and the \keyword{initial data} for solving the constraints as the initial value problem of \Eqsref{ParabolicEquation} -- \eqref{FinalSystemDiffMom}. The solution is therefore a \emph{physical} initial data set. Since we are particularly interested in the asymptotics of these, we start our discussion by analysing a family of spherically symmetric (single black hole) data sets, which we show to be limits at spatial infinity of more general initial data sets in \Sectionref{Sec:Asymptotics_of_binary_black_hole_data} -- in full analogy to the findings in \cite{Beyer:2017tu}.

\subsection{Kerr-Schild-like spherically symmetric solutions of the constraints}
\label{sec:KSsphsymm}
In this section, we discuss a family of spherically symmetric solutions of the constraints encompassing the single Kerr-Schild Schwarzschild black hole data set introduced at the end of \Sectionref{sec:KSID}. This family is obtained by finding the \emph{general spherically symmetric solution} of  the
 parabolic-hyperbolic formulation of the constraints where only the \keyword{free data} are determined by \Eqref{KerrSchildFreeData}. 
We shall discuss that while all these correspond to slices in a Schwarzschild spacetime with a particular mass (as a consequence of the Birkhoff theorem), almost all of these are not asymptotically flat.

In order to construct this family of spherically symmetric initial data sets now we pick any $M\in\R$ and choose the \emph{free data} for \Eqsref{ParabolicEquation} -- \eqref{FinalSystemDiffMom} by \Eqref{KerrSchildFreeData}. Given these free data, we look for the general spherically symmetric solution of \Eqsref{ParabolicEquation} -- \eqref{FinalSystemDiffMom}. As a consequence of spherical symmetry we impose that $p_{a}=0$ and that $q$ and $A$ only depend on $r$. 
Under these assumptions, \Eqsref{ParabolicEquation} -- \eqref{FinalSystemDiffMom} become
\begin{align}
r\partial_{r}q + q=& 2\, \kappa , 
\label{GeneralSolutions1} \\
2r\partial_{r}A - A=&- \left( 1 + \kappa\, q r^2 + \frac{1}{4}\, r^2q^2 \right)A^3.
\label{GeneralSolutions2}
\end{align}
Since the first equation is independent of the second one, we can easily express its 
  general solution as 
\begin{equation}
q=\frac{2\mathcal{C}}{r}-\frac{4 M}{r^2\lambda(r)},
\label{ExactKerr_q}
\end{equation}
where $\mathcal C$ is an integration constant, and we've defined $\lambda(r)$ as in \Eqref{Eq:Def_Lambda}. 
Substituting \Eqref{ExactKerr_q} into \Eqref{GeneralSolutions2} leads to 
\begin{equation}
A= \frac{r\lambda(r)} {\sqrt{r\lambda(r)^2(2(M-m)+\mathcal{C}^{2}r) +  r^2 - 4 M r \lambda(r) \mathcal{C} }},
\label{ExactKerr_A}
\end{equation}
with $m\in\R$ being another integration constant. 
For later reference we note that
\begin{equation}
  \label{eq:sphA}
  A=\frac{1}{\sqrt{\mathcal C^2+1}}+\frac{2\mathcal C M+m}{\left(\mathcal C^2+1\right)^{3/2}r}+O\left(\frac 1{r^2}\right).
\end{equation}
The corresponding initial  data set is then given by
\begin{align}
  \gamma_{ab}&=A^2 dr_a dr_b+h_{ab},\\
  \label{eq:resKabsph}
  K_{ab}&=\kappa A^2 dr_{a}dr_{b}+\frac{1}{2}q h_{ab},
\end{align}
where $A$ is given by \Eqref{ExactKerr_A}, $q$ by \Eqref{ExactKerr_q} and $h_{ab}$ and $\kappa$ by \Eqref{KerrSchildFreeData}.
We see that \eqref{ExactKerr_A} and \eqref{ExactKerr_q}  agree with the particular solution \eqref{KerrSchildSolutions} if and only if $\mathcal C=0$ and $m=M$, in which case, $M$ is the ADM mass.

In order to analyse the asymptotics of this family of data sets, we try to bring $\gamma_{ab}$ to the form \Eqref{FlatLimit2} by introducing a new radial coordinate $R$. It turns out that this is not possible in general. It \emph{is} always possible however to bring the metric asymptotically to the cone form
\begin{equation}
  \psi^2(R)\left\{dR^2 + \left( 1 + \mathcal{C}^2 \right)R^2 \left( d\theta^2 + \sin^{2}\theta\, d\phi^2 \right)\right\}+\ldots,\label{eq:4}
\end{equation}
with
\begin{equation}
  R=\exp\left({\sqrt{1+\mathcal C^2}}\int \frac{A(r)}{r} dr\right)
  =r
  -\frac{2\mathcal{C} M +m}{1 + \mathcal{C}^2}
+O\left(\frac 1{r}\right),
\end{equation}
where
\begin{equation}
  \psi^2(R)=1 + \frac{2 (2\mathcal C M + m)}{(1 + \mathcal C^2)R}+O\left(\frac 1{R^2}\right).
\end{equation}
From the asymptotic form \eqref{eq:4} of the metric one can see that the area of the 2-spheres of constant $R$ is $4\pi R^2(1+\mathcal{C}^2)\ge 4\pi R^2$. Therefore, $\mathcal{C}$ determines the excess area beyond the Euclidean value $4\pi R^2$. If this was an deficit area then this would be the metric of a cone embedded in a 4-dimensional Euclidean space. Nevertheless, we call $\mathcal C$ the \emph{cone parameter}. The metric is asymptotically Euclidean, and hence is of the form \Eqref{FlatLimit2}, if and only if $\mathcal C=0$. Since
\[
  \kappa=\frac{2M}{R^2}+O\left(\frac 1{R^3}\right),
  \quad 
  q=\frac{2\mathcal C}{R} - \frac{4 \left(2\mathcal C^2+1\right) M + 2\mathcal C
   m}{\left(\mathcal C^2+1\right) R^2}+O\left(\frac 1{R^3}\right),\quad p_a=0,
\]
the data set is asymptotically flat with ADM mass $m$ if and only if $\mathcal C=0$. Surprisingly we notice that the mass parameter $M$, which determines the \emph{free data} via \Eqref{KerrSchildFreeData}, has nothing to do with the actual mass of the data set. 

The \textit{Hawking mass} (see \cite{Alcubierre:Book}, Appendix~A) of any $r=const$-sphere
\begin{equation}
\label{HawkingMass}
m_{H}=\sqrt{\frac{| S_r |}{16 \pi}}\left( 1
  +\frac{1}{16\pi} \oint_{S_r} \Theta^{(+)}\Theta^{(-)}
  d S \right),
\end{equation}
where $|{S}_r|$ is the surface area of the $r=\mathrm{const}$-sphere $S_r$, and, where  
\begin{equation}
\Theta^{(\pm)}=-\left( q \pm A^{-1} \kstar \right)
\end{equation}
are the in- and outgoing null expansion scalars defined with respect
to suitably normalised future-pointing null normals of ${S}_r$, turns out to be
\begin{equation}
  \label{eq:sphsymmHawkingmass}
  m_H=m,
\end{equation}
where $m$ is the integration constant found in \eqref{ExactKerr_A}. So, even if $\mathcal C\not=0$ and the initial data set is therefore not asymptotically flat, we can still associate the mass $m$ with this data set.

Thus we find the result that all the spherically symmetric data sets, asymptotically flat or not, have a well-defined and finite Hawking mass limit. This is consistent with the following observation made in \cite{Beyer:2017tu}: It is a consequence of the Birkhoff theorem that for any  spherically symmetric  data set $(\Sigma,\gamma_{ab},K_{ab})$ (which satisfies the constraints) there is a hypersurface in a Schwarzschild spacetime and a diffeomorphism from (a subset of) $\Sigma$ to (a subset of) that hypersurface such that the pull-backs of the first and second fundamental forms induced on that hypersurface by the Schwarzschild metric agree with $\gamma_{ab}$ and $K_{ab}$ (see also \cite{Murchadha:2005ib}). Indeed,  this applies to the asymptotic region of all the spherically symmetric data sets above. Irrespective of the value of $\mathcal C$, the mass of that ``target Schwarzschild spacetime'' turns out to be $m$ as in \Eqref{eq:sphsymmHawkingmass}. This embedding into the target Schwarzschild spacetime is described by the formula
\begin{align}
  t(r)=-r \frac{\mathcal C}{\sqrt{\mathcal C^2+1}}+t_0-
  \left(2 m+\frac{ \mathcal C \left(2\mathcal C^2+3\right) m-2
	M}{\left(\mathcal C^2+1\right)^{3/2}}\right)\ln r +O\left(\frac 1{r}\right),
\end{align}
where $t$ is the time coordinate of the target Schwarzschild spacetime with mass $m$ given in Kerr-Schild form by \Eqsref{GeneralKerr} and \eqref{eq:SSKS} (where $M$ must be replaced by $m$) and where the radial coordinate of $\Sigma$ is mapped to the target radial Schwarzschild Kerr-Schild coordinate (therefore bearing the same name). The quantity $t_0$ is an arbitrary constant.

\subsection{Multiple black hole initial data sets}
\label{sec:constrIDsets}

\subsubsection{Step~1: Multiple black hole data sets}
\label{SubSec:ConstructingAuxiliaryData}

As outlined above, the idea of our initial data construction procedure is to first construct preliminary multiple black hole data sets in a first step without imposing the constraints. The idea is to make  `natural' choices for the fields $l_a$, $V$ and $\dot\gamma_{ab}$ above such that the  data set obtained from the formulas in \Sectionref{sec:KSID} resembles a multiple black hole system at a moment of time. Our particular approach for this is inspired by the work in \cite{Bishop:1998cb}.

To this end, we pick $n$ black hole mass parameters $M_1,\ldots, M_n$ and $n$ Cartesian coordinate position vectors $(x_1,y_1,z_1),\ldots, (x_n,y_n,z_n)$ and set 
\begin{equation}
u=\sum^{n}_{i=1}\frac{M_{i}}{r_{i}},
\label{Potential}
\end{equation}
with
\begin{equation}
  \label{eq:defri}
  r_i= \sqrt{(x-x_i)^2+(y-y_i)^2+(z-z_i)^2}.
\end{equation}
We set
\begin{equation}
  \label{eq:defrho}
  \rho=\frac1u{\sum^{n}_{i=1}M_{i}}
\end{equation}
and notice that $\rho$ is a radial coordinate with 
\begin{equation}
  \label{eq:rhoasymp}
  \rho=r+O(1),\quad r=\sqrt{x^2+y^2+z^2},
\end{equation}
in the limit $r\rightarrow\infty$.

From now on we restrict to the case $n=2$. We write $M_+=M_1$, $M_-=M_2$ and set \[r_+=r_1=(0,0,Z),\quad r_-=r_2=(0,0,-Z)\] for some fixed $Z\ge 0$ and $M_+,M_-\ge 0$. This yields
\begin{equation}
  \label{eq:uBinary}
  u=\frac{M_+}{r_+}+\frac{M_-}{r_-}
\end{equation}
with 
\begin{equation}
  \label{eq:rpmBinary}
  r_\pm=\sqrt{x^2+y^2+(z\pm Z)^2}.
\end{equation}

\begin{figure}[t]
	\centering
	\includegraphics[width=0.95\linewidth]{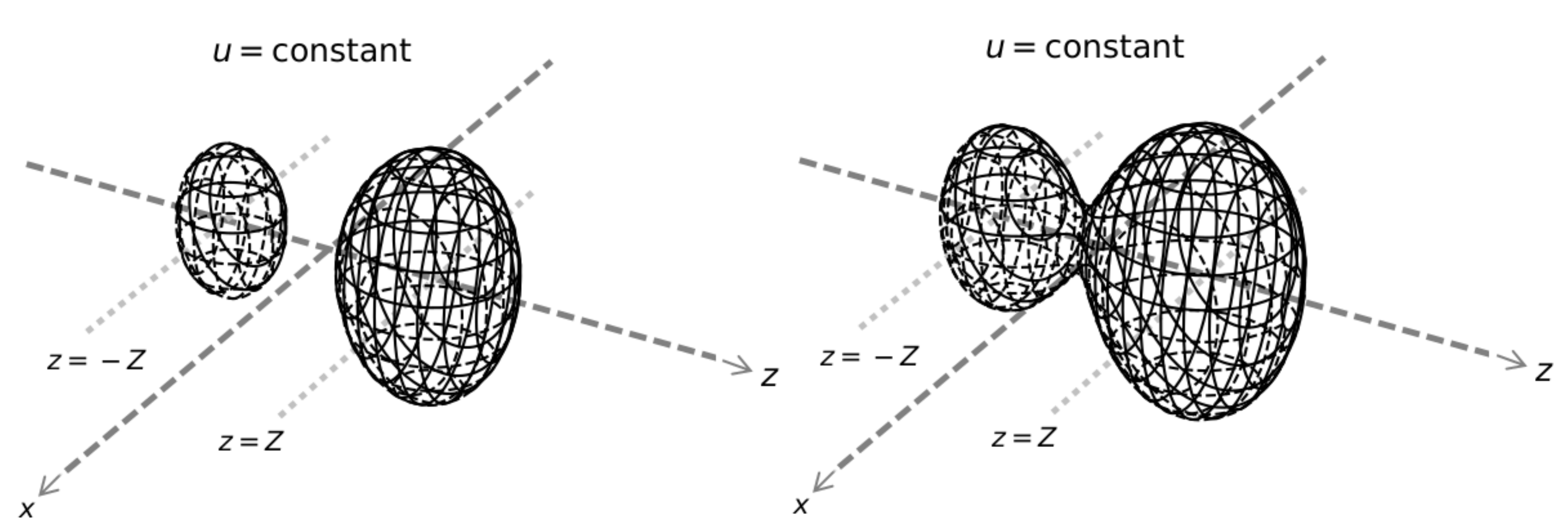}
	\caption{This figure depicts two $u=\text{const}$- (or, equivalently $\rho=\text{const}$-surfaces) as given by \Eqsref{eq:defrho}, \eqref{eq:uBinary} and \eqref{eq:rpmBinary} with $M_{+}>M_{-}$, embedded into $\mathbb{R}^{3}$. The figure on the left shows the surface obtained by a relatively small constant value of $\rho$ which yields two disconnected  spheres. The figure on the right shows the surface obtained by a sufficiently large constant value of $\rho$, creating a single connected peanut-shaped surface diffeomorphic to a single $2$-sphere. Cf.\ \Eqref{eq:peanuttopologychange}.}
	\label{fig:peanut}
\end{figure}
In \Figref{fig:peanut} we show the level-surface $\rho=\rho_0$ for two different values of $\rho_0$. Obviously, these surfaces undergo a topology change as $\rho_0$ varies. Determining the critical value shows that when\footnote{The existence of this lower bound suggests that our current approach to construct data sets is somewhat limited as we cannot put the initial $2$-surface arbitrarily close to the black holes. It is clear however that only small modifications of the geometry of these peanuts is necessary to overcome this problem.}
\begin{equation}
  \label{eq:peanuttopologychange}
\rho_0> \frac{2Z}{\left(\sqrt{M_+}+\sqrt{M_-}\right)^2},
\end{equation} 
the $\rho=\rho_0$-surface is diffeomorphic to a single $2$-sphere; because of the particular shape of these $2$-spheres we sometimes refer to them as \emph{peanuts}. The set of all $\rho=\rho_0$-surfaces satisfying \Eqref{eq:peanuttopologychange} indeed yields a foliation of the exterior region of $\R^3$.

With $\rho$ as defined by \Eqsref{eq:defrho}, \eqref{eq:uBinary} and \eqref{eq:rpmBinary}, we define $l_a$ as in \Eqref{eq:specla} with the positive sign and  $\normall$ given by \Eqref{eq:lanorm}, i.e.,
\begin{equation}
  l_a=\normall\nabla_a\rho,\quad \normall=\frac 1{\sqrt{(\delta^{-1})^{ab}\nabla_a\rho\nabla_b\rho}}
  =\frac 1{\sqrt{\left(\frac{\partial\rho}{\partial r}\right)^2+ \left(\frac{\partial\rho}{\partial\theta}\right)^2}}.
\end{equation}
This means that $l^a$ is outward pointing.
 In order to determine a preliminary Kerr-Schild data set now using the formulas in \Sectionref{sec:KSID}, we pick 
 \begin{equation}
   V=-2u,
 \end{equation}
and
\begin{equation}
  \label{eq:dotgammachoice}
  \dot\gamma_{ab}=\mathcal{F}h_{ab}, 
\end{equation}
for an, at this point, arbitrary function $\mathcal{F}$. It is clear that the resulting data set equals the standard single Kerr-Schild Schwarzschild black hole initial data set discussed in \Sectionref{sec:KSID} if  $M_+=0$ (or $M_-=0$) and $\mathcal F=0$ (with a coordinate shift given by $Z$), or, if $Z=0$ and $\mathcal F=0$. In general notice that by the particular choice of $\dot\gamma_{ab}$ in \Eqref{eq:dotgammachoice}, which we justify in \Sectionref{Sec:Asymptotic_expansions} and \Sectionref{sec:iterations}, only the quantity $q$ is affected by the function $\mathcal F$, see \Eqsref{eq:KSkappa}, \eqref{eq:KSpa} and \eqref{eq:KSqab}.

For arbitrary fixed values of $M_{+},M_{-}$ and $Z$
we can calculate $\kstar$ as
\begin{align}
\kstar = -\frac{2}{\rho}-\frac{2 M_{+}M_{-}Z^2 \left( 1 + 3\cos \left( 2\theta \right) \right)}{ \left( M_{+}+ M_{-} \right)\rho^{3} }+ O\left( \rho^{-4} \right).
\end{align}
It follows that there exists a $\rho_1>0$ such that $\kstar$ is strictly negative (and hence the parabolicity condition \Eqref{eq:parabolcond} holds) for all $\rho>\rho_1$. The fact that we have not been able find an explicit formula for $\rho_1$ is not problem in practice because we can check numerically that $\kstar$ is negative on the whole numerical domain.

It is a consequence of \Eqref{eq:rhoasymp} that
\[
  V=-2(M_++M_-)/r+O(r^{-2})
\]
and hence that each preliminary data set approaches the physical single Kerr-Schild Schwarzschild black hole data set with mass $M_++M_-$ in leading order (provided $\mathcal F$ vanishes zero sufficiently fast in the limit $r\rightarrow\infty$). 

From now on we shall decorate the quantities associated with the  \emph{preliminary data sets} with $\prescript{[P]}{}{}$ in order to distinguish them from corresponding quantities of the final \emph{physical initial data} set discussed in the next subsection.

\subsubsection{Step~2: Solving the parabolic-hyperbolic constraints system}\label{Sec:Numerical_implementation_and_tests}

\paragraph{Adapted coordinates.}
Given a data set as constructed in \Sectionref{SubSec:ConstructingAuxiliaryData}, we next attempt to impose the constraints. To this end, we can read off the free data and the initial data for solving \Eqsref{ParabolicEquation} -- \eqref{FinalSystemDiffMom} as an initial value problem in the increasing $\rho$-direction starting from some $\rho=\rho_0$-surface compatible with \eqref{eq:peanuttopologychange}. For this we  introduce adapted coordinates $(\rho,\vartheta,\varphi)$ where  $\rho$ is the radial coordinate along which we perform the evolutions and where each ``peanut'' $\rho=\text{const}$ is endowed with intrinsic polar coordinates $(\vartheta,\varphi)$. 
The coordinate transformation from the ``original'' spherical coordinates $(r,\theta,\phi)$ 
to these adapted coordinates $(\rho,\vartheta,\varphi)$ is taken to be of the form
\begin{equation}
  \label{eq:coordtrafo}
  \rho(r,\theta,\phi)=(M_++M_-)/u(r,\theta,\phi),\quad
  \vartheta(r,\theta,\phi)=\theta,\quad
  \varphi(r,\theta,\phi)=\phi,
\end{equation}
with $u(r,\theta,\phi)$ given by \Eqsref{eq:uBinary} and \eqref{eq:rpmBinary} using
\[x=r\sin\theta\cos\phi,\quad y=r\sin\theta\sin\phi,\quad z=r\cos\theta.\]

It is clear that while  \Eqsref{ParabolicEquation} -- \eqref{FinalSystemDiffMom} are solved in the adapted $(\rho,\vartheta,\varphi)$-coordinate system, the coefficients of these equations must be calculated from the preliminary data set which has been found  in the original $(r,\theta,\phi)$-coordinate system before. At each $\rho$-step of the numerical evolution, i.e., on any $\rho=\rho_0=\text{const}$-surface, we must therefore compute  $r$  as a function of $\vartheta=\theta$. Since there is no explicit formula for this we find this function\footnote{We shall now mostly suppress the coordinate $\varphi=\phi$ because all examples considered in this paper are axisymmetric and therefore independent of this coordinate.} $\hat r(\vartheta)=r(\rho,\vartheta)_{|\rho=\rho_0}$ numerically as follows\footnote{This is certainly only one way to determine this function numerically. We have not compared this to any other method (for example, the Newton method) yet.}:
\begin{enumerate}
\item Calculate
  \begin{equation}
    \label{eq:hatr0}
\hat r(0)=\frac{1}{2}\left(\left(M_+ + M_-\right)\rho_0+\sqrt{ \left(M_+ + M_-\right)^{2}\rho_0^{2}-4\left( \rho_0\left(M_+ - M_-  \right)Z-Z^2 \right) }  \right),
\end{equation}
which follows by inverting the equation $\rho=\rho_0$ in the special case $\theta=\vartheta=0$.
\item Then numerically solve the ODE
\begin{equation}
\partial_{\vartheta}\hat r(\vartheta)
=\frac{Z\left( M_{+}r_{-}^{3}(\vartheta)
-M_{-}r_{+}^{3} (\vartheta) \right)\hat r(\vartheta)\sin\vartheta}
{M_{-}r_{+}^{3} (\vartheta)\left(\hat r(\vartheta)-Z\cos\vartheta\right)
+M_{+}r_{-}^{3} (\vartheta)\left(\hat r(\vartheta)+Z\cos\vartheta\right)}
\label{r_dot_ODE}
\end{equation}
as an initial value problem on the interval $\vartheta\in [0,\pi ]$ with the initial datum $\hat r(0)$ given by \Eqref{eq:hatr0}.
\end{enumerate}
With the function $\hat r(\vartheta)$ determined in this way for each $\rho_0$ we proceed as follows. The tensor components of all quantities calculated in the $(r,\theta,\phi)$-coordinate system are transformed to the $(\rho,\vartheta,\varphi)$-coordinates using the Jacobi matrix of the coordinate transformation \eqref{eq:coordtrafo} which reads
\begin{equation}
  J(r,\theta,\phi)=
  \begin{pmatrix}
    \partial \rho/\partial r &\partial \rho/\partial\theta & 0\\
    0 &1 & 0\\
    0 &0 & 1
  \end{pmatrix}.
\end{equation}
Observe that its inverse is
\begin{equation}
  J^{-1}(r,\theta,\phi)=\frac{1}{\partial \rho/\partial r}\begin{pmatrix}
    1 & -\partial \rho/\partial\theta & 0\\
    0 &\partial \rho/\partial r & 0\\
    0 &0 & 1
  \end{pmatrix}.
\end{equation}
With $\hat r(\vartheta)=r(\rho,\vartheta)_{|\rho=\rho_0}$ determined on any $\rho=\rho_0=\text{const}$-surface as discussed above, we can express all tensor components  completely in terms of the adapted $(\rho,\vartheta,\varphi)$-coordinates as needed  to numerically solve the constraint equations.

\paragraph{Numerical method, errors and tests.}
At the end of \Sectionref{Sec:ThevacuumConstraintsasaParabolic-HyperbolicSystem}, we have discussed that we can use the \keyword{spin-weight formalism} (see also \Sectionref{Sec:SWSHstuff} in the appendix\footnote{The parabolic hyperbolic equations written in terms of the SWSH can be found in \cite{winicourRacz2018}}) to express all fields intrinsic to the $2$-spheres $\rho=\text{const}$, and thereby the constraint equations \eqref{ParabolicEquation} -- \eqref{FinalSystemDiffMom}, using \keyword{spin-weighted spherical harmonics}; the details of our numerical implementation can be found in \cite{Beyer:2015bv,Beyer:2014bu,Beyer:2016fc,Beyer:2017jw,Beyer:2017tu}. Thus, the ``spatial'' discretisation used in our code is of (pseudo-)spectral nature. All examples considered so far assume axial symmetry, so there is no dependence on $\varphi$. We can therefore employ uniform $\vartheta$-grids with $N$ points (see in particular \cite{Beyer:2016fc} regarding  details) and exploit axial symmetry.
As the ``time''-stepping method we choose the adaptive SciPy ODE solver \textit{\textbf{odeint}}\footnote{See \url{https://docs.scipy.org/doc/scipy/reference/generated/scipy.integrate.odeint.html}.}.
We denote its absolute error tolerance parameter by $\tilde{\mathcal{E}}$ and the corresponding magnitude by $\varepsilon=-\log(\tilde{\mathcal{E}})$\footnote{For this whole paper, $\log$ is the logarithm to base $10$.}. The parameter $\varepsilon$ therefore controls the local $\rho$-step size of the numerical evolutions.

We can expect that the parameters $N$ and $\varepsilon$ can be used to control the error that is numerically generated through the time- and space-discretisation. However, in certain settings we find that the error depends neither on $\varepsilon$ nor on $N$. Then we conclude that the error is dominated by the accumulated finite number representation errors in our code. In such a situation, our code provides only limited means to improve the numerical accuracy. We provide evidence for this in the following sections. 

We now discuss how $\varepsilon$ and $N$ may be used to control the errors. Let $E[f](\varepsilon,N,\rho,\vartheta)$ be the absolute error of some particular unknown $f$ at $(\rho,\vartheta)$ calculated numerically with discretisation parameters $\varepsilon$ and $N$. In principle, this error can of course only be calculated if the exact solution is known. In practice, when the exact solution is not known, we shall follow the common practice  to determine $E[f](\varepsilon,N,\rho,\vartheta)$ by comparing the numerical solution to another numerical solution obtained with some sufficiently high resolution (instead of the exact solution). 

On the one hand, we expect that if $N$ is sufficiently large so that the grid resolves all the spatial features of the solution, the numerical error is dominated by the time discretisation. In such a setting, the numerical error should not become smaller when we increase $N$ (in fact, oversampling may be a significant error source). The error should decrease monotonically with $\varepsilon$. Unless stated otherwise we always pick $\varepsilon=12$. On the other hand, the error can be expected to be dominated by the spatial discretisation if $\varepsilon$ is sufficiently large. In this setting, the error should be roughly independent of $\varepsilon$, but should decrease monotonically with $N$. Unless stated otherwise we always take $N=11$.

In regards to the calculation of the error, we set for any fixed $N$ and $\varepsilon$ 
\begin{equation}
\mathcal{E}_{\rho}(f)=\max_{\vartheta} E[f](\varepsilon,N,\rho,\vartheta),
\end{equation}
which can be used to study the error in both the time and space discretisations.  

\begin{figure}
 	\centering
 	\includegraphics[width=0.99\linewidth]{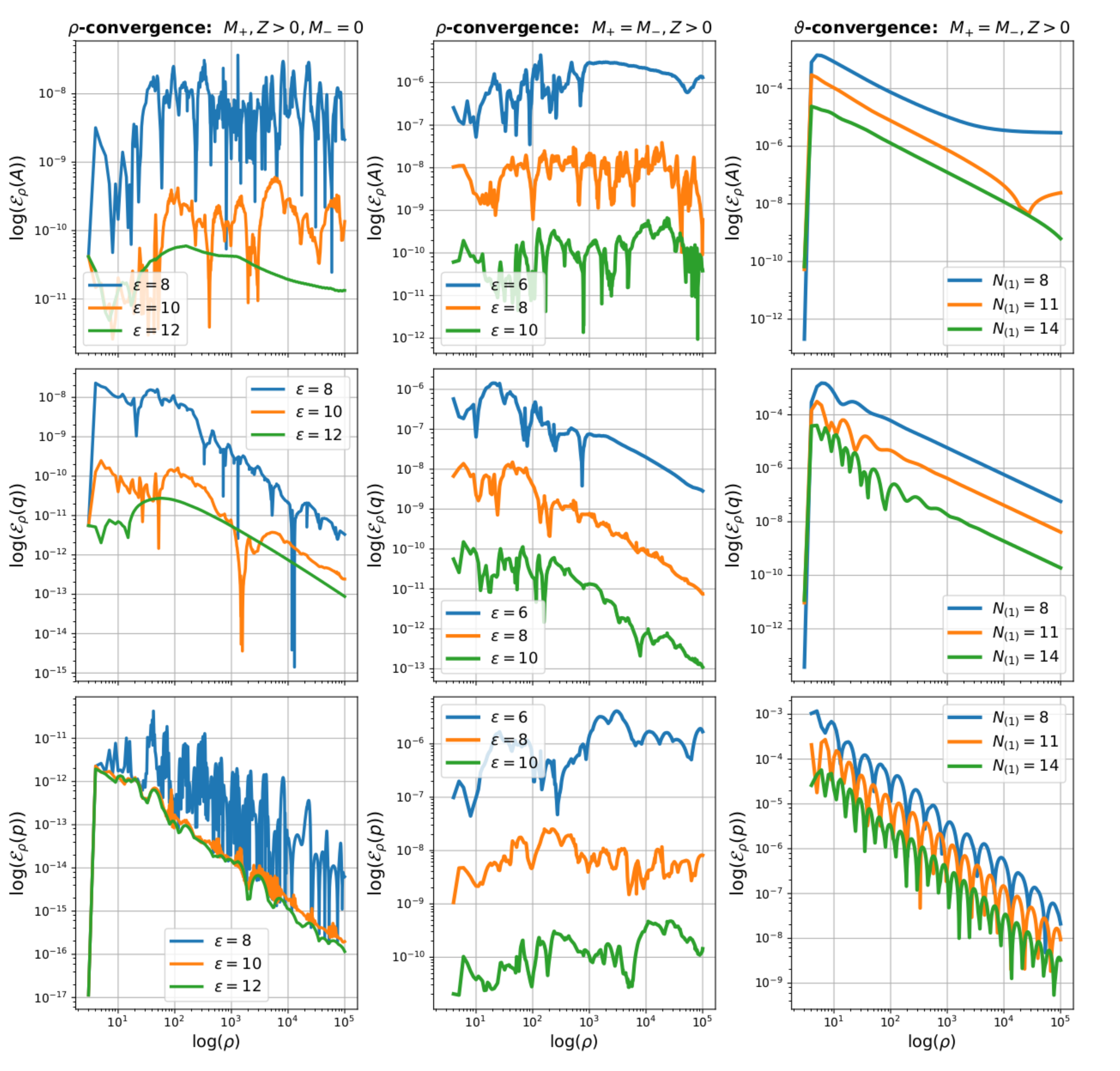}
 	\caption{Numerical convergence tests of our code as discussed in the text at the end of \Sectionref{Sec:Numerical_implementation_and_tests}. The first column shows the shifted single Schwarzschild black hole test case. The second and third columns show the binary black hole test case given by $M_{+}=M_{-}$, $Z=1$ and $\mathcal{F}=0$. }
 	\label{fig:convergence}
 \end{figure}

We now discuss two tests of our code. First we consider a non-trivial test of our numerical implementation by choosing\footnote{In all examples in this paper we will choose $M_++M_-=1$. Any quantity which carries either a distance, time or mass unit is therefore expressed in units of $M_++M_-$ in the geometric physical unit system adopted in this paper. }
$M_{+}=Z=1$ and $M_{-}=\mathcal{F}=0$. Picking $Z>0$ `\textit{shifts}' the single black hole so that it is no longer centred at the origin of the $(r,\theta,\phi)$-coordinate system. The solutions obtained in the adapted coordinates $(\rho,\vartheta,\varphi)$ must however agree with the standard single black hole solution after undoing the shift.  We pick $\varepsilon=8,10,12$ and calculate the quantity $\mathcal{E}_{\rho}$ using the exact single black hole solution as the reference solution. The results are shown in the left column of \Figref{fig:convergence}. The convergence of the functions $q$ and $A$ is consistent with that of our numerical scheme. The error associated with $p$ and $\bar{p}$ 
defined as\footnote{Notice that $p$ has spin-weight $1$ and its complex conjugate $\bar p$ the spin-weight $-1$.}
\begin{equation}
  \label{eq:pppppp}
  p=\frac 1{\sqrt 2} p_a \left(\partial_{\vartheta}^a-{\text{i}}\, {\csc\theta}\, \partial_\varphi^a\right),\quad
  \bar p=\frac 1{\sqrt 2} p_a \left(\partial_{\vartheta}^a+{\text{i}}\, {\csc\theta}\, \partial_\varphi^a\right),
\end{equation}
are below the numerical round-off error which occurs at $10^{-15}$. 

The second test case is given by the choice $M_{+}=M_{-}=1/2$, $Z=1$ and $\mathcal{F}=0$. No exact solution is known here and the numerical solution obtained with some higher resolution is therefore chosen as the reference solution to calculate the error. The right most column of \Figref{fig:convergence} shows convergence plots for the spatial discretisation. When a numerical solution is calculated with a spatial resolution $N_{(1)}$, the reference solution is calculated with a spatial resolution  $N_{(2)}=2N_{(1)}-1$ to ensure that both numerical resolutions share grid points. The calculated errors shown in both the middle and right column of \Figref{fig:convergence} have the expected dependence on $\varepsilon$ and $N$.

%%%%%%%%%%%%%%%%%%%%%%%%%%%%%%%%%%%%%%%%%%%%%%%%%%%%%%%%%%%%%%%%%%%%%%%%%%%%%%%%%%%%%%%%%%%%%%%%%%
%-------------------------------------------------------------------------------------------------
\section{Asymptotic properties}
\label{Sec:Asymptotics_of_binary_black_hole_data}

\subsection{Formal asymptotic expansions and their numerical justifications}
\label{Sec:Asymptotic_expansions}

In \Sectionref{Sec:Binary_black_hole_model} we have proposed a construction procedure for a class of initial data sets and discussed some numerical test cases. The hope is that these initial data sets describe binary black hole systems. Whether this is really the case remains to be seen. In this section we make an important step towards the physical understanding of these data sets by analysing  the asymptotics $\rho\rightarrow\infty$. 

To this end first recall the results in the spherically symmetric case in \Sectionref{sec:KSsphsymm}. We found there that every spherically symmetric data set is uniquely determined by the parameters $\mathcal{C}$, $m$ and $M$, and is asymptotically flat with ADM mass $m$ if and only if the cone parameter $\mathcal{C}$ vanishes. In any case, the limit of the Hawking mass at $\rho\rightarrow\infty$ is always $m$ irrespective of the value of $\mathcal C$.
Intuitively, we would expect that \emph{any}  initial data set obtained as in \Sectionref{Sec:Binary_black_hole_model} should become spherically symmetric asymptotically. The asymptotics found in the spherically symmetric case should therefore apply to this larger family of initial data sets as well.

In order to provide evidence for this claim, we perform a formal power series analysis. To this end, we first calculate power series expansions (with respect to $\rho$ in the limit $\rho\rightarrow\infty$) of the general class of binary black hole preliminary data sets in \Sectionref{SubSec:ConstructingAuxiliaryData}. Recall that this yields
the free data for the constraint equations \eqref{ParabolicEquation} -- \eqref{FinalSystemDiffMom}. Second, we make the following power series ansatz for the unknowns $q$, $A$ and $p_a$ of \Eqsref{ParabolicEquation} -- \eqref{FinalSystemDiffMom}\footnote{Consistently with footnote~\ref{ftn:OSymbol1}, the $O$-symbol is defined component-wise with respect to a natural coordinate system in the case of tensor quantities  (here, the $(\rho,\vartheta,\varphi)$-coordinates).}:
\begin{align}
\label{eq:Apowerseries}
A(\rho,\vartheta,\varphi)
&={A^{(0)}(\vartheta,\varphi)}+\frac{A^{(1)}(\vartheta,\varphi)}{\rho}+O\left( \frac{1}{\rho^{2}}\right),
\\
\label{eq:qpowerseries}
q(\rho,\vartheta,\varphi)
&=\frac{q^{(1)}(\vartheta,\varphi)}{\rho}+\frac{q^{(2)}(\vartheta,\varphi)}{\rho^2}+O\left( \frac{1}{\rho^{3}}\right),
\\
\label{eq:ppowerseries}
p_a(\rho,\vartheta,\varphi)
&=\frac{p_{a}^{(1)}(\vartheta,\varphi)}{\rho}+O\left( \frac{1}{\rho^2} \right).
\end{align}
The structure of the equations and the expressions of the free data suggest that it is sufficient to consider integer powers of $\rho$ only\footnote{This is significantly different for the results in \cite{Beyer:2017tu} obtained with the hyperbolic formulation of the constraints where also half-integer powers of $\rho$ are present.}. Shortly we will present numerical support for this ansatz. If these expansion do not fairly represent the asymptotic behaviour of the solutions then we would see evidence of this in our numerics.  

Before we proceed we notice that all the following expansions are independent of the choice of $\mathcal{F}$ in \Eqref{eq:dotgammachoice}. In fact it follows from the discussion in \Sectionref{SubSec:ConstructingAuxiliaryData} and \Eqsref{eq:gendotgamma} -- \eqref{eq:KSqab} for the particular choice \Eqref{eq:dotgammachoice} that only the function $\prescript{[P]}{}{q}$, i.e., the function $q$ associated with the \emph{preliminary} data set in Step~1 of our construction procedure  in \Sectionref{sec:constrIDsets}, is affected by the choice of $\mathcal F$. This consequently only affects the initial data of the resulting function $q$ in Step~2 of our construction procedure. While these initial data certainly affect  the resulting \emph{values} of the expansion coefficients in the following, the expansions themselves and the particular relationships between the various expansion coefficients, which we uncover now, hold irrespectively.

It turns out that the leading term of the expansion of \Eqref{FinalSystemDiffMom}, i.e., the term of order $\rho^{-1}$, yields the equation 
\begin{equation*}
\hat{D}_{a}\left( \frac{q^{(1)}}{A^{(0)}}\right) =0\implies \frac{q^{(1)}}{A^{(0)}}=\text{constant}.
\end{equation*}
Recall that $\hat{D}_{a}$ is the covariant derivative associated with the metric $\Omega_{ab}$ of the standard round unit sphere. Motivated by the spherically symmetric case, see \Eqsref{ExactKerr_q} and \eqref{eq:sphA}, we write 
\begin{equation}
  \label{eq:formexpq1A0}
  q^{(1)}(\vartheta,\varphi)=2{\mathcal C \sqrt{\mathcal C^2+1}}\, A^{(0)}(\vartheta,\varphi)
\end{equation}
for some (possibly new) constant $\mathcal C$.
Then, the leading term of the expansion of \Eqref{ParabolicEquation}, i.e., the term of order $\rho^{-2}$, turns out to yield the nonlinear elliptic equation
\begin{equation}
\label{eq:A0eq}
A^{(0)}\hat\Delta A^{(0)}={A^{(0)}{}^2(1+\mathcal C^2(\mathcal C^2+1) A^{(0)}{}^2)} -1 =: A^{(0)} F[A^{(0)}],
\end{equation}
where $\hat\Delta = \Omega^{ab} \hat D_a \hat D_b$ is the Laplacian on the round unit sphere. 
First we notice that if $A^{(0)}$ is any smooth solution of this equation, then $A^{(0)}$ does not change sign because $A^{(0)}(\vartheta,\varphi)=0$ at any point $(\vartheta,\varphi)$ would clearly violate \Eqref{eq:A0eq} at that point. We therefore conclude that any smooth solution of this equation is either strictly positive or strictly negative. Suppose we have a smooth strictly \emph{negative} solution $A^{(0)}$. Then $-A^{(0)}$ is a smooth strictly positive solution. Assuming that $A^{(0)}$ is strictly positive from now on is therefore no loss of generality.  A particular smooth strictly positive solution is the \textit{constant} solution \begin{equation} \label{eq:LA0sol} A^{(0)}=\frac{1}{\sqrt{\mathcal C^2+1}}; \end{equation} compare this to \Eqref{eq:sphA}.  In fact, \Eqref{eq:LA0sol} is the only smooth strictly positive (constant or not) solution of \Eqref{eq:A0eq}: Suppose there were two different smooth strictly positive solutions $A^{(0)}$ and $\tilde A^{(0)}$ of \Eqref{eq:A0eq}. Then a standard integration by parts argument implies
\begin{equation}
  \label{eqLA0eq2} -\|\hat D (A^{(0)}-\tilde A^{(0)})\|=\left<A^{(0)}-\tilde A^{(0)},F[A^{(0)}]-F[\tilde A^{(0)}]\right>,
\end{equation}
where the norm and the scalar product here are the standard $L^2$-norm and $L^2$-scalar product on the $2$-sphere with respect to $\Omega_{ab}$. One can easily check that
\[
  F[A^{(0)}]-F[\tilde A^{(0)}]
=\frac{{A^{(0)}} \tilde A^{(0)}\left(\left(\mathcal{C}^2+1\right) \mathcal{C}^2
\left({A^{(0)}}^2+{A^{(0)}} {\tilde A^{(0)}}+\tilde {A^{(0)}}^2\right)+1\right)+1}{
   {A^{(0)}} \tilde A^{(0)}}(A^{(0)}-\tilde A^{(0)}).
\]
Since the fraction on the right-hand side is strictly positive if $A^{(0)}$ and $\tilde A^{(0)}$ are strictly positive, the right-hand side of \Eqref{eqLA0eq2} is therefore non-negative. But since the left-hand side is non-positive, $A^{(0)}$ and $\tilde A^{(0)}$ must be equal. Thus, we conclude that $A^{(0)}$ given by \Eqref{eq:LA0sol} is the only strictly positive smooth solution of \Eqref{eq:A0eq}. Due to \Eqref{eq:formexpq1A0} we  therefore have
\begin{equation}
  \label{eq:q1sol}
  q^{(1)}=2\mathcal C
\end{equation}
in agreement with the spherically symmetric case \Eqref{ExactKerr_q}. Due to \Eqsref{eq:LA0sol}, \eqref{eq:Apowerseries} and the leading behaviour of $h_{ab}$ we can now interpret $\mathcal C$ again as a cone parameter just as in the spherically symmetric case. In particular it follows that \emph{the resulting initial data set is not asymptotically flat if $\mathcal C\not=0$}.

Again in agreement with the spherically symmetric case, the Hawking mass  \Eqref{HawkingMass} always has a well-defined limit $m$ as a function of $\rho$ if \Eqsref{eq:Apowerseries} -- \eqref{eq:ppowerseries} hold irrespective of whether $\mathcal C$ is zero or not. In fact,
\begin{equation}
\label{eq:hawkinglimitgen}
m_{H}(\rho)
= \underbrace{ \underline{A^{(1)}} \left({\mathcal C}^2+1\right)^{3/2}-2\,{\mathcal C}
  (M_++M_-)}_{=:m} + O\left( \frac{1}{\rho}\right), 
\end{equation}
where $\underline{A^{(1)}}$ is the average of the function $\underline{A^{(1)}}$ over the round unit sphere\footnote{According to \Eqref{ec:mean_value_s2} and \Eqref{eq:functionS2axial}, we have $\underline f=f_0/\sqrt{4\pi}$ if $f$ is any axially symmetric smooth function on $\mathbb S^2$.}, see \Eqref{eq:average}, and where we have used that both the  $\rho^{-3}$-order of \Eqref{ParabolicEquation} and the  $\rho^{-3}$-order of \Eqref{FinalSystemDiffNorm} imply that
\[
  \underline{q^{(2)}}=-4 (M_++M_-).
\]

We have therefore found that if \Eqsref{eq:Apowerseries} -- \eqref{eq:ppowerseries} hold then the general solution of the constraint equations \eqref{ParabolicEquation} -- \eqref{FinalSystemDiffMom} with free data determined by the preliminary data sets in \Sectionref{SubSec:ConstructingAuxiliaryData} is spherically symmetric in leading order and the limit of the Hawking mass is consistent with the spherically symmetric case. 

In general, however, the solutions are not spherically symmetric beyond the leading order.
We have seen that the initial data set is not asymptotically flat and the $3$-metric is not asymptotically Euclidean if $\mathcal C\not =0$. In the case $\mathcal C=0$,  the next order of the expansion of the constraint equations takes the form
\begin{equation}
  \label{eq:A1eqasldk}
\hat\Delta A^{(1)}=0, \quad \hat\Delta q^{(2)}=-8 (M_+ + M_-) - 2 q^{(2)},\quad p^{(1)}_a=\frac 12\hat D_a q^{(2)}.
\end{equation}
This implies that
\[A^{(1)}(\vartheta,\varphi)=\underline{A^{(1)}}=const,\]
and the general axially symmetric solution for $q^{(2)}$ is 
\begin{equation}
  \label{eq:q2euclidean}
  q^{(2)}(\vartheta,\varphi)=-4 (M_+ + M_-)+q^{(2)}_1\, {}_{0}Y_{1}(\vartheta),
\end{equation}
where $q^{(2)}_1$ is any real constant and where ${}_{0}Y_{1}$ is a spherical harmonic (see \Sectionref{Sec:SWSHstuff}). In particular, we have the important equivalence
\begin{equation}\label{Eq:ImportantEquivalence}
p^{(1)}_a=0\quad\Longleftrightarrow\quad q^{(2)}_1=0.
\end{equation}
We can therefore conclude that the resulting \textit{initial data set is asymptotically flat (Definition~\ref{AsympFlatDef}) if and only if $\mathcal C=0$ and  $q^{(2)}_1=0$}. While the condition $\mathcal C=0$ alone implies that the $3$-metric is asymptotically Euclidean, the  initial data set is only asymptotically flat if in addition $q^{(2)}_1=0$. Notice that according to \cite{bartnik1986} the ADM mass is therefore only uniquely defined if both 
$\mathcal C=0$ and  $q^{(2)}_1=0$. It is remarkable however that the limit of the Hawking mass \Eqref{eq:hawkinglimitgen} with respect to our foliation of $2$-spheres  always exists. Whether this limit is unique or whether it depends on our particular foliation of $2$-spheres remains open.

\begin{figure}[t]
	\centering
	\includegraphics[width=0.95\linewidth]{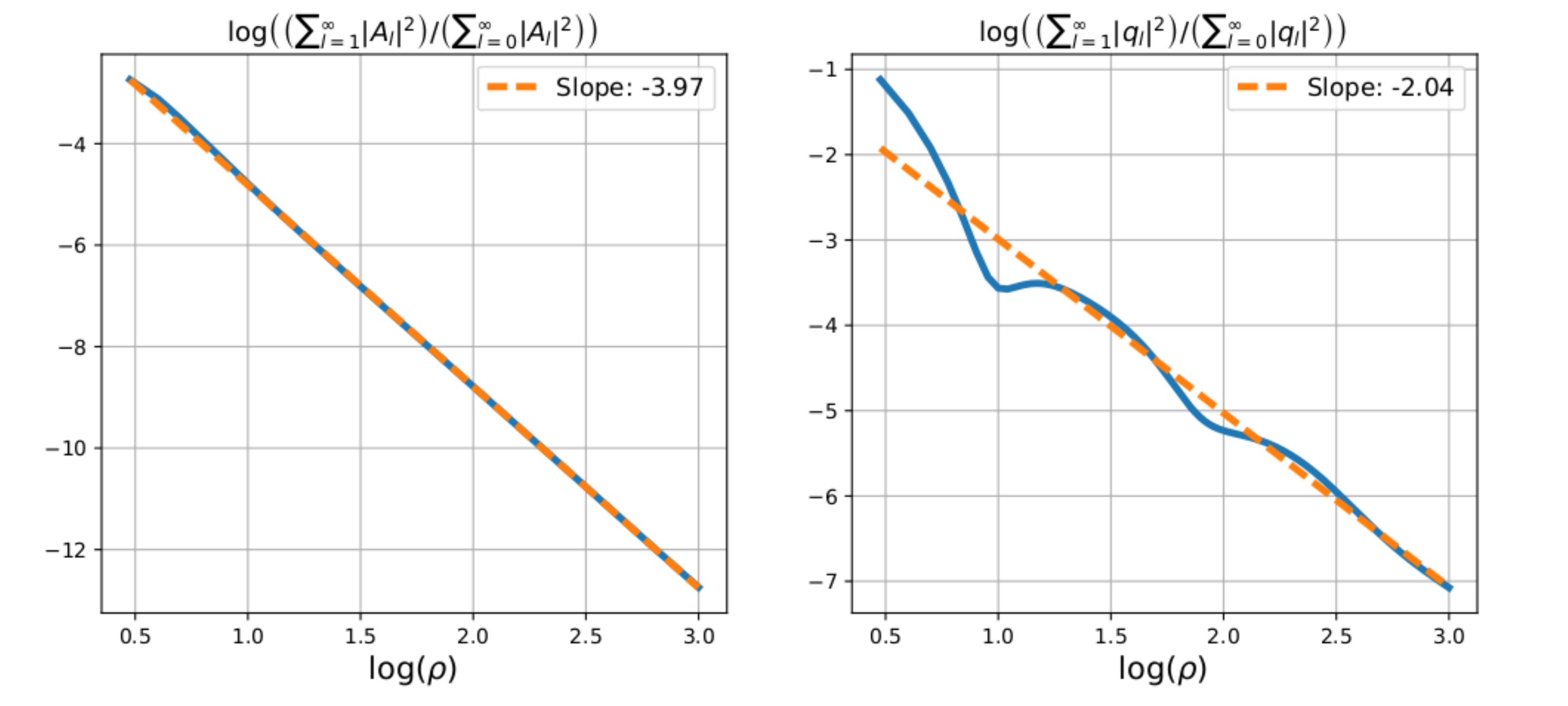}
	\caption{Numerical decay properties of the functions $A$ (left) and $q$ (right) for $\rho\to\infty$ in the case  $M_{-}=2/3$, $M_{+}=1/3$ and $Z=1$. }
	\label{fig:decaytest}
\end{figure}

In order to support this formal asymptotic analysis we show numerical results for  $M_{-}=2/3$, $M_{+}=1/3$ and $Z=1$ in \Figref{fig:decaytest}. In order to check our prediction that $A\rightarrow\underline A$ in leading-order in the limit $\rho\rightarrow\infty$, the first plot in \Figref{fig:decaytest} shows the quantity
\[1
-\frac{4\pi |\underline A(\rho)|^2}{\|A(\rho)\|^2_{L^2(\mathbb{S}^2)}}
=\frac{\|A(\rho)\|^2_{L^2(\mathbb{S}^2)}-4\pi |\underline A(\rho)|^2}{\|A(\rho)\|^2_{L^2(\mathbb{S}^2)}}=\frac{\sum_{l=1}^\infty |A_{l}(\rho)|^2}{\sum_{l=0}^\infty |A_{l}(\rho)|^2},\]
where we have used \Eqsref{ec:mean_value_s2}, \eqref{eq:L2} and \eqref{eq:functionS2axial}. Our results suggest that this quantity should be $O(\rho^{-2})$ for large $\rho$. Our numerical findings are certainly consistent with this.
Next we check our prediction that $q\rightarrow\underline q$ in leading-order in the limit $\rho\rightarrow\infty$ by plotting the quantity
\[1
-\frac{4\pi |\underline q(\rho)|^2}{\|q(\rho)\|^2_{L^2(\mathbb{S}^2)}}
=\frac{\sum_{l=1}^\infty |q_{l}(\rho)|^2}{\sum_{l=0}^\infty |q_{l}(\rho)|^2}\]
in \Figref{fig:decaytest}. Again we find agreement with this prediction.

\begin{figure}[t]
	\centering
	\includegraphics[width=0.65\linewidth]{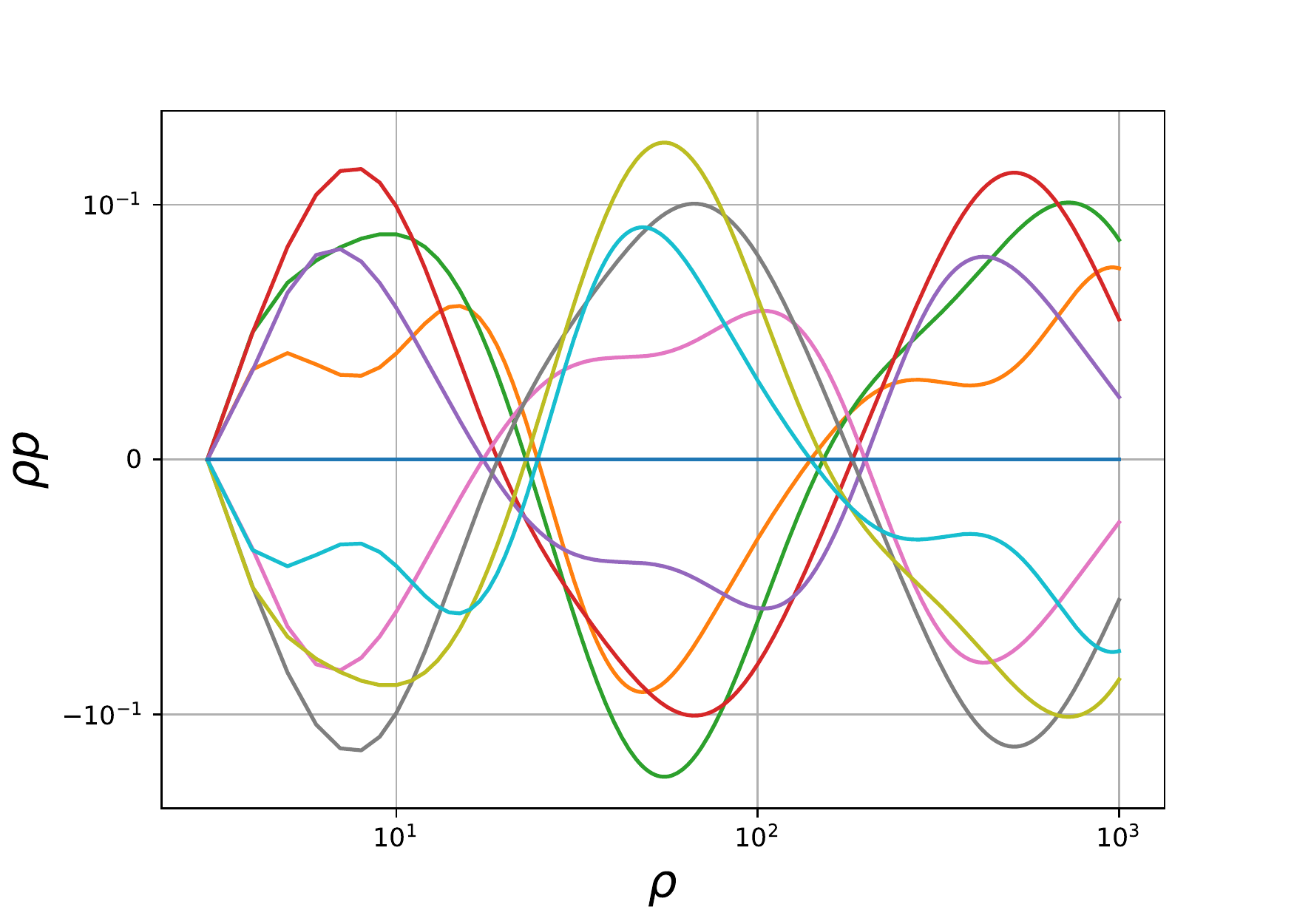}
	\caption{A \textit{`symlog`} plot \cite{webber2012} showing the numerical values of $\rho p(\rho,\vartheta)$ for different values of $\vartheta$ in the case $M_+=1/2$, $M_-=1/2$ and $Z=1$.}
	\label{fig:spheretestp}
\end{figure}

Finally, we also provide numerical evidence to support \Eqref{eq:ppowerseries} in\footnote{This plot uses  the ‘symlog’ option in matplotlib in Python, which for a function $f(x)$ essentially plots $\text{sign}(f(x)) \log (|f(x)|)$, for details see for instance \cite{webber2012}.} Fig.~\ref{fig:spheretestp} where we plot $\rho p$ as a function of $\rho$ demonstrating that it remains finite for large values of $\rho$. Recall that based on \Eqsref{eq:A1eqasldk} and \eqref{eq:q2euclidean} we do not expect $p_{a}^{(1)}$, and therefore $q^{(2)}_1$, to vanish even in the asymptotically Euclidean case $\mathcal C=0$.

\subsection{Numerical determination of the asymptotic parameters}
\label{sec:numasymptoticparameters}

Given any initial data set as in \Sectionref{Sec:Binary_black_hole_model}, how would we calculate the asymptotic parameters identified in \Sectionref{Sec:Asymptotic_expansions} numerically?

\begin{figure}[t]
	\centering
	\includegraphics[width=1.\linewidth]{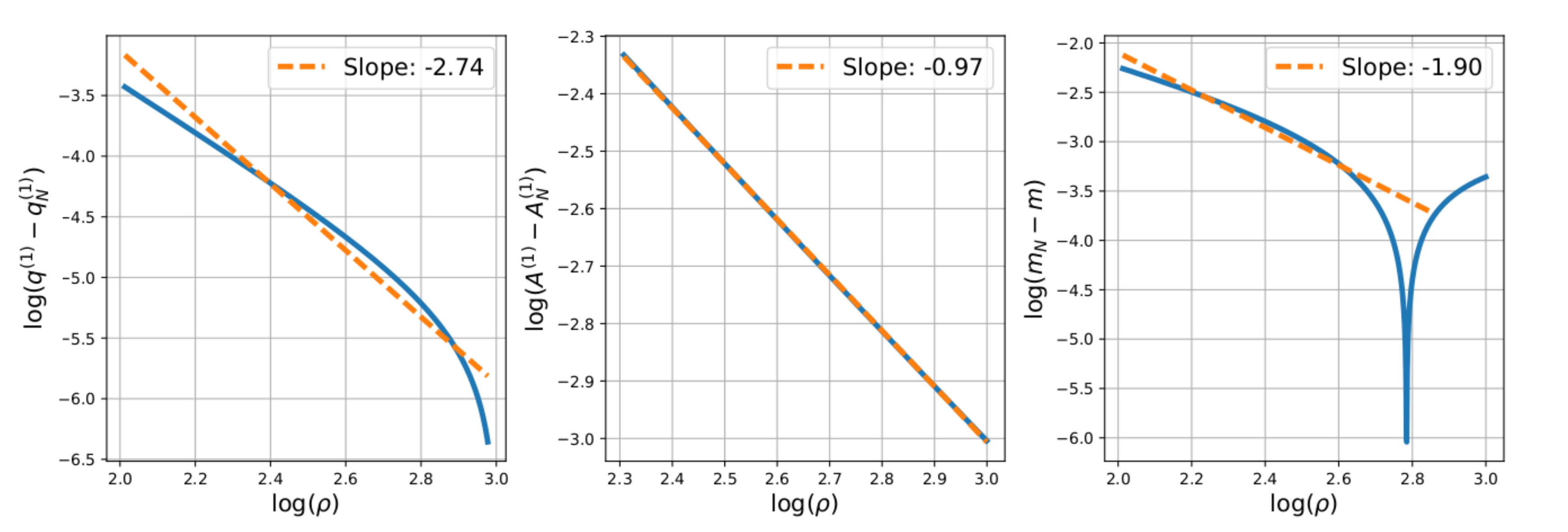}
	\caption{Numerical support for the formal analytical predictions in \Eqsref{eq:Cconvrate}, \eqref{eq:Aconvrate} and \eqref{eq:mconvrate}. This evolution was carried out with $M_- = 2/3$, $M_+=1/3$ and $Z=1$ where we find $q^{(1)}=4.5\times 10^{-4},\;\; \underline {A^{(1)}}=0.932$ and $m = 0.94$.}
	\label{fig:matchingdecay}
\end{figure}

The results in \Sectionref{Sec:Asymptotic_expansions} suggest that
\[\underline{q}(\rho)=\frac{2\mathcal C}{\rho}-\frac{4(M_++M_-)}{\rho^2}+O(\rho^{-3}),\]
and $\underline q$ therefore agrees with the spherically symmetric case in the two leading orders. Calculating $\underline{q}(\rho)$ for some sufficiently large $\rho$ therefore allows us to numerically estimate the cone parameter $\mathcal C$ as follows. We write
\begin{equation*}
  \mathcal C=\frac{1}{2}\rho\underline{q}(\rho)+\frac{2(M_++M_-)}{\rho}+O(\rho^{-2}).
\end{equation*}
Defining
\begin{equation}
  \label{eq:CNrho}
  \mathcal C_N(\rho)=\frac{1}{2}\rho\underline{q}(\rho)+\frac{2(M_++M_-)}{\rho},
\end{equation}
it follows that $\mathcal C_N(\rho)\rightarrow \mathcal C$ in the limit $\rho\rightarrow\infty$, in fact,
\begin{equation}
  \label{eq:Cconvrate}
  \mathcal C=\mathcal C_N(\rho)+O(\rho^{-2}).
\end{equation}
So $\mathcal C_N(\rho)$ in \Eqref{eq:CNrho} can be understood as a numerical approximation of the asymptotic parameter $\mathcal C$ which converges to $\mathcal C$ in the limit $\rho\rightarrow\infty$ with the rate $O(\rho^{-2})$. This is confirmed in the first plot of \Figref{fig:matchingdecay}.

In a similar way the above results yield
\[\underline A(\rho)=\frac{1}{\sqrt{\mathcal C^2+1}}+\frac{\underline {A^{(1)}}}{\rho}+O(\rho^{-2}).\]
Hence,
\[\underline {A^{(1)}}=\rho\left(\underline A(\rho)-\frac{1}{\sqrt{\mathcal C^2+1}}\right)+O(\rho^{-1}).\]
In analogy to the above, the quantity
\begin{equation}
  \underline {A^{(1)}}_N(\rho)=\rho\left(\underline A(\rho)-\frac{1}{\sqrt{\mathcal C_N^2(\rho)+1}}\right)
\end{equation}
is therefore a numerical approximation of $\underline {A^{(1)}}$, in fact,
\begin{equation}
  \label{eq:Aconvrate}
  \underline {A^{(1)}}=\underline {A^{(1)}}_N(\rho) +O(\rho^{-1})
\end{equation}
using that  
\[\frac{1}{\sqrt{\mathcal C^2+1}}=\frac{1}{\sqrt{\mathcal C_N^2(\rho)+1}}+O(\rho^{-2})\]
as a consequence of \Eqref{eq:Cconvrate}. \Eqref{eq:Aconvrate} is  confirmed in the second plot of \Figref{fig:matchingdecay}.
This can be used to estimate the limit $m$ of the Hawking mass in \Eqref{eq:hawkinglimitgen}. We have
\begin{align*}
m&=\underline{A^{(1)}} \left({\mathcal C}^2+1\right)^{3/2}-2\,{\mathcal C}
   (M_++M_-)\\
&=\underline{A^{(1)}}_N(\rho) \left({\mathcal C}^2 _N(\rho)+1\right)^{3/2}-2\,{\mathcal C}_N(\rho)
   (M_++M_-)+O(\rho^{-1}).
\end{align*}
The quantity
\begin{equation}
  m_N(\rho)=\underline{A^{(1)}}_N(\rho) \left({\mathcal C}^2 _N(\rho)+1\right)^{3/2}-2\,{\mathcal C}_N(\rho)
   (M_++M_-)
\end{equation}
is therefore a numerical approximation of $m$ and
\begin{equation}
  \label{eq:mconvrate}
  m=m_N(\rho)+O(\rho^{-1}).
\end{equation}
This decay is confirmed in the third plot in \Figref{fig:matchingdecay}

Even though the estimates $\mathcal C_N(\rho)$, $m_N(\rho)$ for $\mathcal C$, $m$ etc.\ are better the larger the value of $\rho$ is at which we calculate $\mathcal C_N(\rho)$, $m_N(\rho)$, we find that numerical errors in numerically solving the constraint equations become significant when we go further than $\rho\sim 10^{3}$ (for $\epsilon=12$ and $N=11$; see the end of \Sectionref{Sec:Numerical_implementation_and_tests}). The question is therefore how good the values $\mathcal C_N(\rho)$ and $m_N(\rho)$ at, say, $\rho\sim 10^{3}$ are as approximations for the actual asymptotic parameters $\mathcal C$ and $m$. For any of these quantities $\mu$ and the corresponding function $\mu_N(\rho)$, we consider 
\[\mathcal E_A[\mu]=|\mu_N(2\rho)-\mu_N(\rho)|\]
calculated at some large value of $\rho$ as a measure for the absolute error.
Consider $M_+ = M_-=\frac 12$ and $Z=1$ as an example. Our numerical evolution is carried out for $\rho_{0}=3$. 
Our estimation of the asymptotic parameters at $\rho=10^3$ yields
\begin{align}
\mathcal{C}=0.0774,\; \mathcal{E}_{A}(\mathcal{C})=9.5\times 10^{-7},\;\; m = 0.9399,\;\; \mathcal{E}_{A}(m)=7.86\times 10^{-7}.
\end{align}
Notice that the relative error is of order $\sim 10^{-6}$. As mentioned above, this is likely due to the error of measuring $\mathcal{C}$ and $m$ at a finite point and not at infinity. However, due to the errors in numerically solving the constraints we find that we cannot go further than $\rho\sim 10^3$. So we need to accept whatever error we have at that point in the measurement of the matching parameters.

\subsection{Iterative construction of asymptotically Euclidean binary black hole data}
\label{sec:iterations}

Consistent with our previous work \cite{Beyer:2017tu} we have seen here also that solving the constraints as an initial value problem has the major drawback of giving no  control over the asymptotics of the solutions. With incorrect or unphysical asymptotics however the resulting initial data sets may not have any reasonable physical interpretation. The last example in \Sectionref{sec:numasymptoticparameters} for instance is not asymptotically flat (neither in the weak nor the strong sense) -- a property one would expect for any compact isolated gravitationally bound astrophysical system. In fact it is not even asymptotically Euclidean. The problem is that the relationship between the freely specifiable quantities $M_+$, $M_-$, $Z$ and $\mathcal F$ and the resulting corresponding asymptotic quantities $\mathcal C$, $m$ and $q^{(2)}_1$ is highly nonlinear and nonlocal, and, therefore hard to analyse. As described earlier, this involves first the construction of a preliminary data set for given $M_+$, $M_-$, $Z$ and $\mathcal F$ (i.e., Step~1 in \Sectionref{SubSec:ConstructingAuxiliaryData}), second, the numerical evolution (Step~2 in \Sectionref{Sec:Numerical_implementation_and_tests}) for as large values of $\rho$ as possible, in order to determine the asymptotic parameters as in \Sectionref{sec:numasymptoticparameters}. 

Now we want to address the following question: Suppose that for some choice of $M_+$, $M_-$, $Z$ and $\mathcal F$ the corresponding asymptotic parameters found as described above are not ``favourable'', in particular consider the case that $\mathcal C$ does not vanish. How can we change the free quantities $M_+$, $M_-$, $Z$ and $\mathcal F$ in order to ``improve the  data set''? In fact we shall now discuss an iterative numerical procedure which allows us to decrease $|\mathcal C|$ step by step and thereby, in principle, make $|\mathcal C|$ as small as we like. We shall see that the function $\mathcal F$ introduced in \Eqref{eq:dotgammachoice} is crucial for this. 
Recall our discussion in \Sectionref{Sec:Asymptotic_expansions} which yields that  $\mathcal C=0$ is not sufficient for asymptotic flatness because $q^{(2)}_1$ must vanish as well. In this section now we nevertheless restrict our attention to the cone parameter $\mathcal C$. We believe that similar ideas apply to deal with $q^{(2)}_1$ as well. 

\begin{figure}[t]
	\centering
	\includegraphics[width=0.95\linewidth]{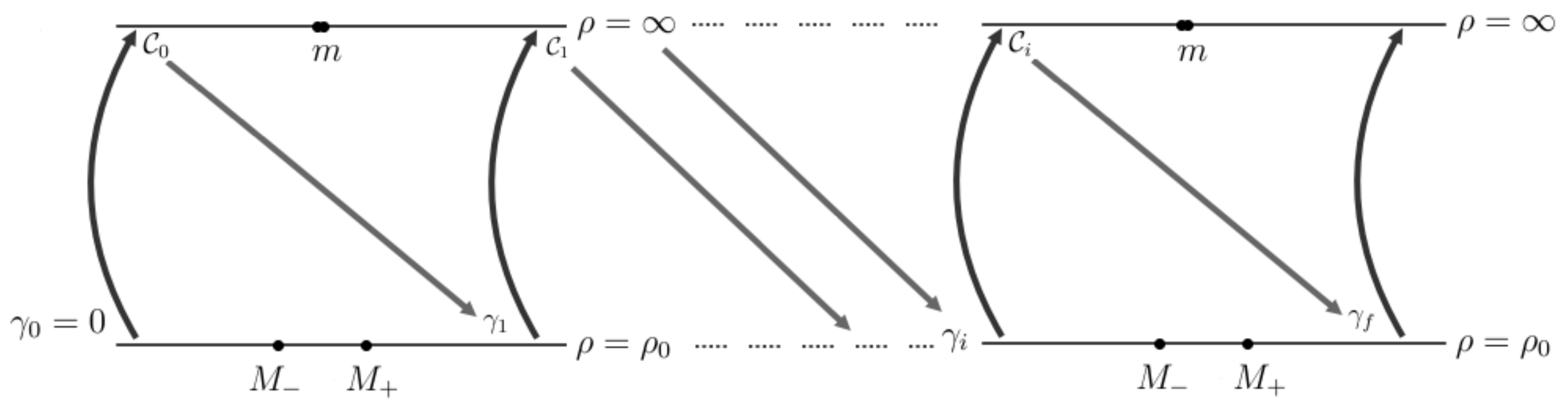}
	\caption{An illustration of the iterative scheme introduced in \Sectionref{sec:iterations}.}
	\label{fig:iterativeillistration}
\end{figure}

The method is based on the following observation. It is a consequence of \Eqsref{eq:KSkappa}, \eqref{eq:KSpa}, \eqref{eq:KSqab}, \eqref{eq:dotgammachoice} and the procedure in \Sectionref{SubSec:ConstructingAuxiliaryData} that the only quantity of the preliminary data set constructed in Step~1 of our procedure which is affected by the function $\mathcal F$ is ${}^{[A]}q$. In fact, if we consider the parameters $M_+$, $M_-$ and $Z$ as fixed now and consider two preliminary data sets given by  two different choices $\hat\gamma$ and $\tilde\gamma$ of a real parameter $\gamma$ introduced by
\[\mathcal F=\frac{2\gamma}{\rho\sqrt{1-V}},\]
\Eqref{eq:KSqab} yields that
\[{}^{[P]}\hat q-{}^{[P]}\tilde q=-\frac{\hat \gamma-\tilde \gamma}{\rho}+O(\rho^{-2}).\]
It is clear that such two different preliminary data sets lead to two different initial data sets in Step~2 of our method in \Sectionref{Sec:Numerical_implementation_and_tests}. According to \Eqsref{eq:qpowerseries} and \eqref{eq:q1sol}, it can be expected that both resulting data sets have two different cone parameters $\hat{\mathcal C}$ and $\tilde{\mathcal C}$ and
\[\hat q-\tilde q=2\frac{\hat{\mathcal C}-\tilde{\mathcal C}}{\rho}+O(\rho^{-2}).\]
At least in certain regimes we could therefore expect that
\begin{equation}
  \label{eq:approxlindepiter}
  \hat{\mathcal C}-\tilde{\mathcal C}=-\frac 1\nu (\hat \gamma-\tilde \gamma)
\end{equation}
for some positive approximately constant quantity\footnote{The  heuristic argument here might suggest that $\nu\approx 1$. The following iteration scheme turns out to converge faster if we pick $\nu$ as in \Eqref{eq:kappaiter} below.} $\nu$. 
Based on this, we propose the following iterative scheme\footnote{We have not yet tried to use any classical root finding method like the bisection or the secant method. It is possible that some of these converge faster or are more reliable than the one discussed here.}. As above, fix the free quantities $M_+$, $M_-$, $Z$, and set
\begin{equation}
  \label{eq:kappaiter}
  \nu=\frac{Z}{M_++M_-}.
\end{equation}
First calculate the full initial data set as in \Sectionref{sec:constrIDsets} for 
 $\gamma_0=0$,
 and determine the corresponding cone parameter $\mathcal C_0$ as in \Sectionref{sec:numasymptoticparameters}. Then pick $\gamma_1=\nu\mathcal C_0$, and, again calculate the full initial data set and determine the cone parameter $\mathcal C_1$. If the (purely experimental) choice of $\nu$ in \Eqref{eq:kappaiter} was ``correct'' and \Eqref{eq:approxlindepiter} was therefore exact, we would find $\mathcal C_1=0$.  We would have therefore achieved the goal of ``improving'' the cone parameter and find an initial data set with the optimal value $\mathcal C_1=0$. In general, however, the resulting value of $\mathcal C_1$ will in general not be zero. Numerical evidence suggests that  $|\mathcal C_1|<|\mathcal C_0|$. We therefore repeat the iteration as often as necessary until eventually the resulting cone parameter is sufficiently close to zero. In summary, our proposed iterative procedure is defined as follows. Fix $M_+$, $M_-$, $Z$ and set $\nu$ as in \Eqref{eq:kappaiter}. Pick an accuracy goal parameter $\mu>0$.
  \begin{description} 
  \item [Start condition:] Pick $\gamma_0=0$. Calculate the corresponding full initial data set as in \Sectionref{sec:constrIDsets} and determine the corresponding cone parameter $\mathcal C_0$ as in \Sectionref{sec:numasymptoticparameters}. 
  \item [Iterative step:] Let $n\ge 0$. Suppose that we have determined $\gamma_n$, the corresponding full initial data set and the corresponding cone parameter $\mathcal C_n$. If $|\mathcal C_n|<\mu$, stop here. Otherwise, set
    \begin{equation}
      \label{eq:iterstepgamma}
      \gamma_{n+1}=\gamma_n+\nu\mathcal C_n,
    \end{equation}
    and repeat the iterative step with $n$ replaced by $n+1$.
  \end{description}
Again notice that the motivation for \Eqref{eq:iterstepgamma} is that if \Eqref{eq:approxlindepiter} was exact with $\nu$ as in \Eqref{eq:kappaiter}, then $\mathcal C_{n+1}=0$. The scheme is illustrated in \Figref{fig:iterativeillistration}.

We shall now apply this to three cases as summarised below. For all these three cases we choose $\rho_0=3$ and estimate the asymptotic parameters at $\rho=10^3$. We always pick $\epsilon=12$, $N=11$ and $\mu=10^{-9}$. Notice that due to the fixed choice $\rho_0=3$, \Eqref{eq:peanuttopologychange} gives a restriction of how much we can vary the parameters $M_+$, $M_-$ and $Z$.
\begin{case}
	\item[$M_{+}=1/2$, $M_{-}=1/2$, $Z=1$:]
	For this case, only two iterations are needed until $\mathcal{C}$ has a sufficiently small magnitude\footnote{We write $\mathcal C=\mathcal C_n$ and $\gamma_f=\gamma_n$ where $n$ is the last iteration step.}: 
	\[
	\mathcal{C}=2.13\times 10^{-9},\, \mathcal{E}_{A}(\mathcal{C})=4.61\times 10^{-9}.
	\]
	The corresponding value of $\gamma_f$ is
	\[
	\gamma_f =1.38\times 10^{-4},\, \mathcal{E}_{A}(\gamma_f)=4.83\times 10^{-10},
	\]
	and the mass is
	\[
	m=9.33\times 10^{-1},\, \mathcal{E}_{A}(m)=4.25\times 10^{-4}.
	\]
	\item[$M_{+}=2/3$, $M_{-}=1/3$, $Z=1$:]
	A total of eight iterations are needed here to give a sufficiently small cone parameter
	\[
	\mathcal{C}=3.19\times 10^{-8},\, \mathcal{E}_{A}(\mathcal{C})=1.34\times 10^{-9},
	\]
	with
	\[
	\gamma_f =3.27\times 10^{-4},\, \mathcal{E}_{A}(\gamma_f )=2.78\times 10^{-9},
	\]
	and mass
	\[
	m=9.35\times 10^{-1},\, \mathcal{E}_{A}(m)=2.09\times 10^{-4}. 
	\]
	\item[$M_{+}=1/2$, $M_{-}=1/2$, $Z=3/2$:]
	Fourteen iterations are used to get the values:
	\[
	\mathcal{C}=3.80\times 10^{-8},\, \mathcal{E}_{A}(\mathcal{C})=1.59\times 10^{-9}.
	\]
	Here
	\[
	\gamma_f =2.34\times 10^{-2},\, \mathcal{E}_{A}(\gamma_f )=4.89\times 10^{-5}.
	\]
        and
	\[
	m=6.15\times 10^{-1},\, \mathcal{E}_{A}(m)=1.05\times 10^{-3}.
	\]
\end{case}

\begin{figure}
	\centering
	\includegraphics[width=0.99\linewidth]{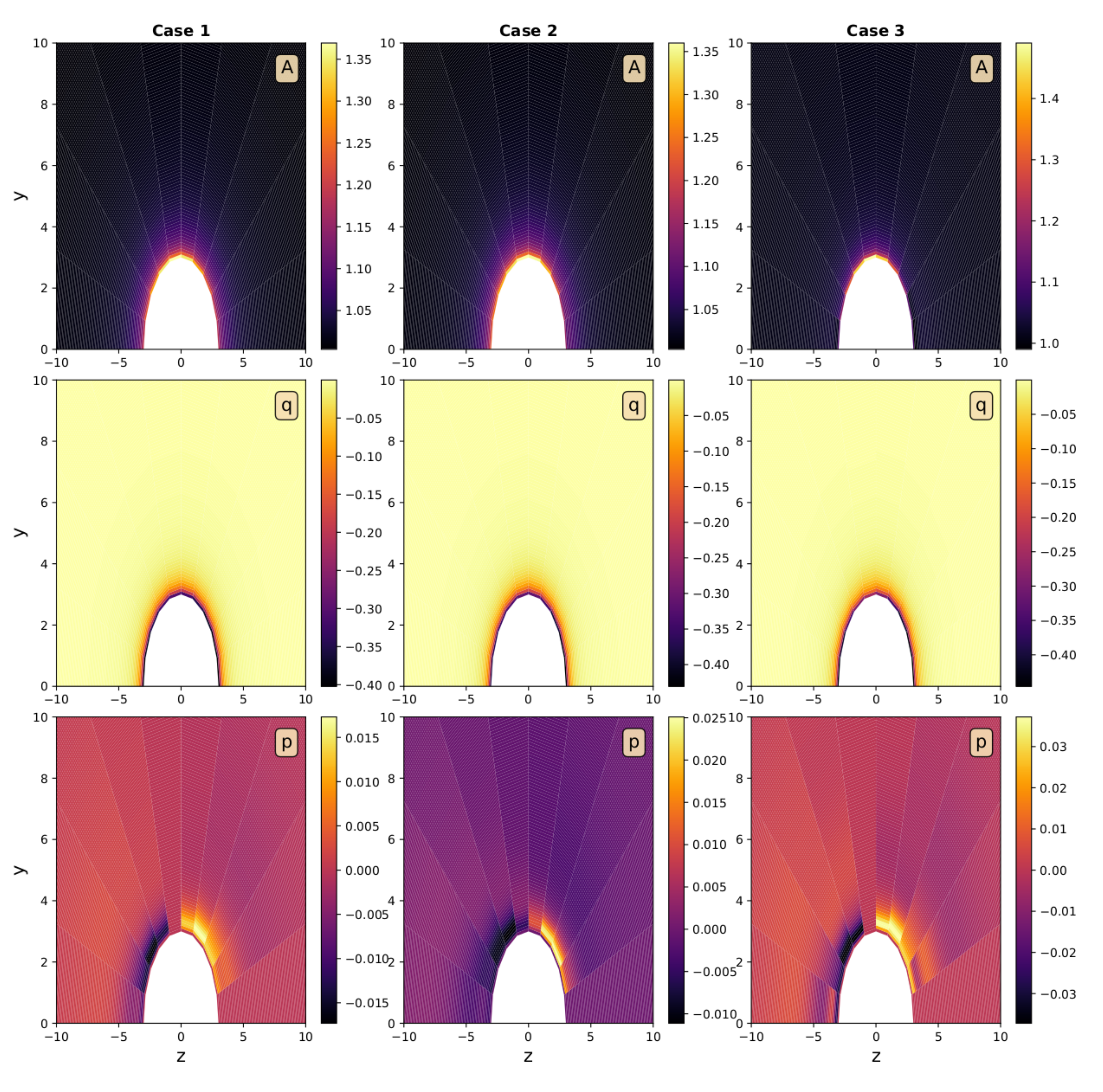}
	\caption{Plots for the three examples considered in \Sectionref{sec:iterations}. Each column presents one of the three cases there. Since the solutions are axially symmetric we only show the $z$-$y$ plane. We have `zoomed' in on the solutions near $\rho_{0}$ as they quickly become homogeneous.  }
	\label{fig:iterationcases}
\end{figure}
 \Figref{fig:iterationcases} shows colour maps of the solutions, for each case, in ``Cartesian'' coordinates.

\begin{equation*}
x=\rho \sin\theta\sin\phi,\;\; y=\rho\sin\theta\cos\phi,\;\; z=\rho\cos\theta.
\end{equation*}

\begin{figure}[t]
	\centering
	\includegraphics[width=0.95\linewidth]{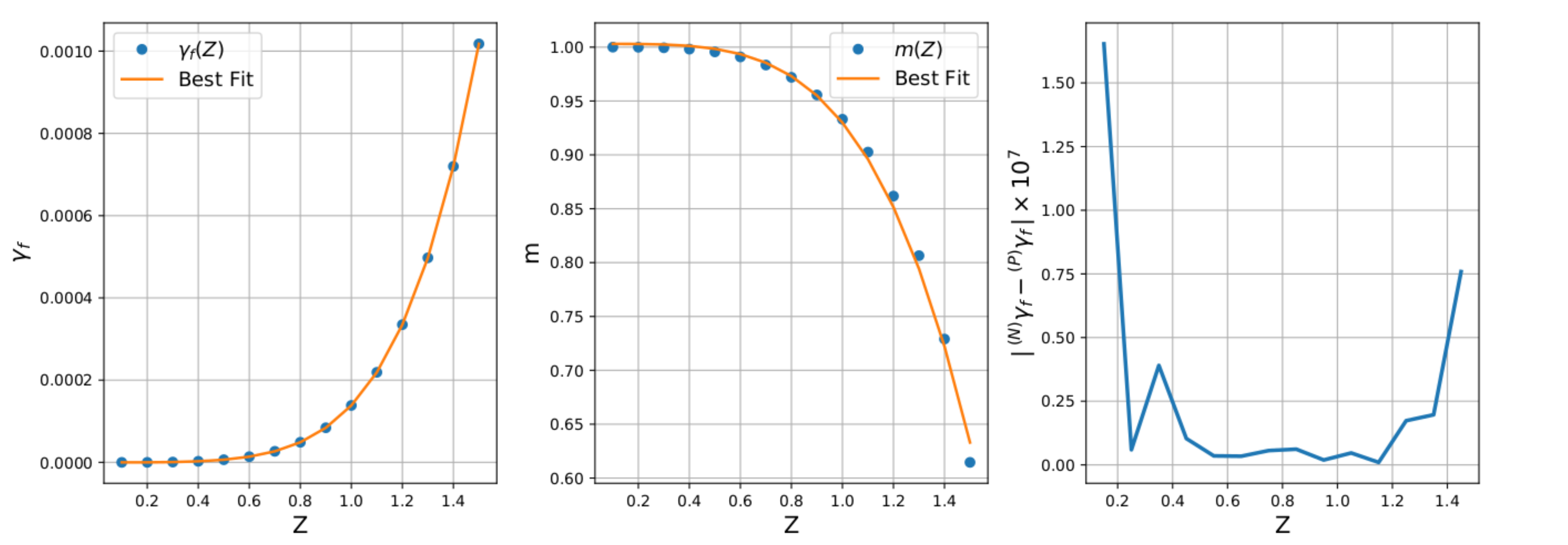}
	\caption{Dependence of $\gamma_f$ and $m$ on $Z$ for solutions of the constraints obtained by the iterative procedure in \Sectionref{sec:iterations}. Here we fix $M_{+}=M_{-}=1/2$. A polynomial of order four is fitted to each of these functions. The right plot shows the difference of the values of $\gamma_f$ determined by the iteration scheme ($\prescript{(N)}{}{\gamma}_f$) and the value given by the fitted polynomial ( $\prescript{(P)}{}{\gamma}_f$).}
	\label{fig:iterationssequence}
\end{figure}

We wish to point out several things. First, it turns out that in none of the three cases above the resulting quantity $q^{(2)}_1$ is zero; the initial data sets are therefore not asymptotically flat. Therefore, one should not regard them as physical. Second, we have noticed that the resulting masses $m$ (the well-defined limits of the Hawking mass) only depend weakly on the mass parameters $M_+$ and $M_-$ as long as $M_++M_-$ is kept constant (here we choose $M_++M_-=1$). The masses $m$, however, depend strongly on $Z$. In order to further study the dependence of $m$ and $\gamma_f$ on $Z$ we fix $M_+ = M_-=1/2$ and solve the equations for various separation distances $Z\in[0.1,1.5]$. Both $\gamma_f$ and $m$ are shown as functions of $Z$ in  \Figref{fig:iterationssequence}.
In the left plot in \Figref{fig:iterationssequence} we see that $\gamma_f$  is an increasing function of $Z$. By fitting a fourth order polynomial to the numerical values of $\gamma_f$ we are able to interpolate this function. The right plot shows the difference of the values of $\gamma_f$ determined by the iteration scheme ($\prescript{(N)}{}{\gamma}_f$) and the value given by the fitted polynomial ( $\prescript{(P)}{}{\gamma}_f$). Regarding the second plot in \Figref{fig:iterationssequence}, it is interesting to notice that $m$ is a decreasing function of $Z$. For $Z=0$, i.e., the single black hole case, we get $m=1$ as expected. When the separation parameter $Z$ is larger, the mass decreases. This is counter-intuitive, particularly when compared to the Newtonian case, where the gravitational binding energy should become small as $Z$ increases. In fact, even within GR it is expected~\cite{Dain:2002cm} that the interaction energy of a binary black hole system behaves like
\[
  E =  - \frac{M_+M_-}{2Z} + O(Z^{-3})
\]
which should increase the total energy $m = M_++M_- +E$ with increasing $Z$. There are several possibilities to interpret this observation: one could simply dismiss this consideration on the basis that the data sets we compute are only asymptotically Euclidean and not asymptotically flat. So they do not provide data for a physical vacuum system but for some system that may have a source at infinity. A second possibility for the behaviour could be that the initial data do not correspond to ``two black holes'' since we have not demonstrated that there are apparent horizons in the data set. Finally, if we accept that these data sets do possess some physical relevance, then it could still be argued that we are not in the regime in which the asymptotic formula holds. Whatever the true answer is, it is clear that this phenomenon requires a better understanding and more numerical work.

We end with a note about the convergence rate of our iterative scheme. For Case 1 (defined by $2M_+ = 2M_- = Z = 1$) it took $126$ iterations to get the cone parameter to a magnitude of $10^{-12}$. This is a marked increase from the two iterations to get it to $\sim 10^{-9}$. Use of more advanced root finding methods such as the Newton method or the bisection method would likely increase the rate of convergence.

\section{Conclusions}\label{Sec:Conclusion}
We construct initial data sets using a particular adaptation of \cite{Bishop:1998cb} to the parabolic-hyperbolic formulation of the constraint equations by R\'acz in a multiple black hole setting.
Similar to \cite{Beyer:2017tu}, where we constructed perturbed single black hole initial data sets with the algebraic-hyperbolic formulation of the constraint equations, we find that our initial data sets here are in general not asymptotically flat. While the initial data sets in our previous work are at least asymptotically Euclidean (because the full $3$-metric can be prescribed freely), the initial data sets here are  asymptotically cone-like unless the asymptotic quantity $\mathcal C$ defined above vanishes. This and the other relevant asymptotic quantity $q^{(2)}_1$ are  determined by the free data and initial data in a complicated non-linear and non-local fashion which is far from being understood. It is therefore impossible to predict which choice of initial data and free data implies $\mathcal C=q^{(2)}_1=0$ which is necessary for asymptotic flatness. Our results suggest that is only the case for a set of initial data sets of measure zero.

Given this we then demonstrate by incorporating some further degrees of freedom (in particular non-vanishing values for the function $\mathcal F$) that asymptotically Euclidean initial data sets can be approximated by an iterative scheme. We have not incorporated the quantity $q^{(2)}_1$ into this scheme yet, but are hopeful that similar ideas can be applied to iteratively approximate asymptotically flat initial data sets in a similar way. This could be done by taking  advantage of  further degrees of freedom present in the choice of $\dot{\gamma}_{ab}$ in \Eqref{eq:dotgammachoice}.

Now, even if one of our initial data set is asymptotically flat, it is not clear whether it represents black holes. The potential for demonstrating the existence of apparent horizons, which would indicate the presence of black holes, is however restricted by the limitation \Eqref{eq:parabolcond} imposed by R\'acz's formalism for foliations based on $2$-spheres. According to this, the constraints must be evolved outwards and can therefore not be used to explore the interior geometry. In contrast to this, the restriction implied by \Eqref{eq:peanuttopologychange} as a consequence of the particular ``peanut shape'' of our $2$-spheres, which prevents us from putting the ``initial'' inner boundary arbitrarily close to the ``black holes'', can be overcome straightforwardly. In any case, for this paper we have not made any efforts to search for (pieces of) apparent horizons in our data sets. The focus of this paper is purely on the asymptotics at spacelike infinity.
We have not investigated other potential remedies for this.  One possibility could be to choose the initial value for the lapse $A$ as negative which may allow us to integrate inwards. Another solution could be to do the inwards-integration from the initial $2$-surface with the \emph{algebraic-hyperbolic formulation} of the constraints, while the  outwards-integration is performed with the \emph{parabolic-hyperbolic formulation} as before.

%%%%%%%%%%%%%%%%%%%%%%%%%%%%%%%%%%%%%%%%%%%%%%%%%%%%%%%%%%%%%%%%%%%%%%%%%%%%%%%%%%%%%%%%%%%%%%%%%%
%-------------------------------------------------------------------------------------------------
\section*{Acknowledgements}

JR was supported by a Master scholarship awarded by the University of Otago. Part of this research was funded by a grant to JF from the Division of Sciences of the University of Otago.

\begin{appendices}
\setcounter{equation}{0} % To restart the counting 
\numberwithin{equation}{section} % To counting in the appendix 
\appendixpage
\appendix

\section{Notation and conventions comparison}
\label{A_NotationConvention} 
In this work we have chosen our sign conventions as in the texts \cite{Alcubierre:Book,Gourgoulhon:Book}. We have also attempted to simplify the notation, which differs from both R\'{a}cz's work and from our previous work \cite{Beyer:2017tu}. The following table provides a comparison of the different conventions: 

\begin{center}
\begin{adjustbox}{width=0.99\textwidth}
\def\arraystretch{1.6}
\begin{tabular}{c|c|c|c|c|c|c|c|c|c|c|c|c|c|c }
	& $K_{ab}$&$\kappa$&$p_{a}$&$Q_{ab}$&$q$&$\gamma_{a b}$&$h_{ab}$&$A$&$B_a$&$N_a$&$v_a$&$k_{ab}$&$\nabla_{a}$&$D_a$  \\ 
	\hline 
	\cite{Beyer:2017tu}&$-\chi_{ab}$&$-\boldsymbol{\kappa}$&$-\boldsymbol{k}_a$&$-\mathring{K}_{ab}$&$-\boldsymbol{K}$&$h_{ab}$&$\hat{\gamma}_{ab}$&$\hat{N}$&$\hat{N}_a$&$\hat{n}_a$&$\dot{\hat{n}}_{a}$&$-\hat{K}_{ab}$&$D_{a}$&$\hat{D}_a$\\ 
	\hline 
	R{\'a}cz&$-\chi_{ab}$&$-\boldsymbol{\kappa}$&$-\boldsymbol{k}_a$&$-\mathring{K}_{ab}$&$-\boldsymbol{K}$&$h_{ab}$&$\hat{\gamma}_{ab}$&$\hat{N}$&$\hat{N}_a$&$\hat{n}_a$&$\dot{\hat{n}}_{a}$&$-\hat{K}_{ab}$&$D_{a}$&$\hat{D}_a$
\end{tabular} 
\end{adjustbox}
\end{center}

\section{Spin-weight and spin-weighted spherical harmonics}\label{Sec:SWSHstuff}
We say that a function $f$ defined on $\mathbb{S}^2$ has \emph{spin-weight $s$} if it transforms as $f \to e^{\text{i} s \xi} f$ under a local rotation by an angle $\xi$ in the tangent plane at any point in $\mathbb{S}^2$. Let $(\vartheta,\varphi)$ be standard polar coordinates on $\mathbb S^2$. If $f$ has spin-weight $s$ and is sufficiently smooth, it can be written as
\begin{equation}\label{eq:functionS2}
f(\vartheta,\varphi)=  \sum\limits_{l=|s|}^{\infty}  \sum\limits_{m=-l}^{l} f_{lm}\, {}_{s}Y_{lm} (\vartheta,\varphi),
\end{equation}
where $_{s}Y_{lm}( \vartheta , \varphi)$ are the \emph{spin-weighted spherical harmonics (SWSH)} and where $f_{lm}$ are complex numbers. Using the conventions in \cite{Penrose:1984tf,Beyer:2015bv,Beyer:2014bu,Beyer:2016fc,Beyer:2017jw,Beyer:2017tu}, these functions satisfy
\begin{equation}\label{integral_properties_spherical_harmonics}
  \int \limits_{\mathbb{S}^2} \  {}_{s} Y_{l_1 m_1 }(\vartheta,\varphi)
  \: _{s}\overline{Y}_{l_2 m_2}(\vartheta,\varphi) \ d\Omega = \delta_{l_1 l_2} \delta_{m_1 m_2},
\end{equation}
where $\delta_{lm}$ is the Kronecker delta and $d\Omega$ is the area element of the metric of the round unit sphere. Using this we find that the coefficients $f_{lm}$ in \Eqref{eq:functionS2} can be calculated as
\begin{equation}
  f_{lm}=\int \limits_{\mathbb{S}^2} f(\vartheta,\varphi)\, {}_{s}\overline{Y}_{lm} (\vartheta,\varphi) d\Omega.
\end{equation}

The \textit{eth-operators} $\eth$ and $\eth'$ are  defined by 
\begin{equation}\label{eq:def_eths}
\eth f       = \partial_\vartheta f - \dfrac{\text{i}}{ \sin \vartheta} \partial_\varphi f- s f \cot \vartheta, \quad 
\eth' f = \partial_\vartheta f + \dfrac{\text{i}}{ \sin \vartheta} \partial_\varphi f + s f \cot \vartheta  ,
\end{equation}
for any function $f$ on $\mathbb{S}^2$ with spin-weight $s$. We have
\begin{align}\label{eq:eths}
\eth  \hspace{0.1cm}_{s}Y_{lm} (\vartheta,\varphi)  &= - \sqrt{ (l-s)(l+s+1) } \hspace{0.1cm}_{s+1}Y_{lm} (\vartheta,\varphi) , \\
\label{eq:eths2}
\eth'   \hspace{0.1cm}_{s}Y_{lm} (\vartheta,\varphi)   &= \sqrt{ (l+s)(l-s+1) } \hspace{0.1cm}_{s-1}Y_{lm} (\vartheta,\varphi) , \\
\eth' \eth  \hspace{0.1cm}_{s}Y_{lm} (\vartheta,\varphi)   &= - (l-s)(l+s+1) \hspace{0.1cm}_{s}Y_{lm} (\vartheta,\varphi) .
\end{align}
Thus, using the properties above it is easy to see that $\eth$ raises the spin-weight by one while $\eth'$ lowers it by one.

In our discussion we are often interested in the \emph{average} of a function $f$ with spin-weight $0$ on $\mathbb{S}^2$ defined by 
\begin{equation}
\label{eq:average}
\underline{f} = \dfrac{1}{4 \pi} \int \limits_{\mathbb{S}^2} \: f d\Omega.
\end{equation}
Expressing $f$ in terms of SWSH and using \Eqref{integral_properties_spherical_harmonics} it follows 
\begin{equation}\label{ec:mean_value_s2}
  \begin{aligned}
\underline{f} &= \dfrac{1}{4 \pi}  \int \limits_{\mathbb{S}^2} \: \sum\limits_{l=0}^{\infty}  \sum\limits_{m=-l}^{l} f_{lm}\, {}_{0}Y_{lm} (\vartheta,\varphi) \; d\Omega , \\
    &= \dfrac{\sqrt{4\pi}}{4 \pi }  \int \limits_{\mathbb{S}^2} \: \sum\limits_{l=0}^{\infty}  \sum\limits_{m=-l}^{l} f_{lm}\,{}_{0}Y_{lm} (\vartheta,\varphi) \; _{0}\overline{Y}_{00}(\vartheta,\varphi) \; d\Omega , \\
    & =   \frac{1}{\sqrt{4\pi}}f_{00},
  \end{aligned}
\end{equation}
where we have used the fact that $_{0}Y_{00}(\vartheta,\varphi) =  (4\pi)^{-1/2}$. 
Another quantity of interest is the $L^2$-norm with respect to the standard round metric on $S^2$. The \emph{Parseval identity} states that
\begin{equation}
  \label{eq:L2}
  \|f\|^2_{L^2(\mathbb{S}^2)}=\sum_{l=0}^\infty\sum_{m=-l}^l |f_{lm}|^2.
\end{equation}

Finally we notice that all functions considered in this paper are axially symmetric and therefore do not depend on the angle $\varphi$. For such functions, all coefficients with $f_{lm}$ with $m\not=0$ vanish and we use the following short-hand notation to write \Eqref{eq:functionS2} as
\begin{equation}\label{eq:functionS2axial}
f(\vartheta)=  \sum\limits_{l=|s|}^{\infty}  f_{l}\, {}_{s}Y_{l} (\vartheta).
\end{equation}

\end{appendices}
%\bibliographystyle{abbrv}
%\bibliography{bibfil}

\begin{thebibliography}{10}

\bibitem{Alcubierre:Book}
M.~Alcubierre.
\newblock {\em Introduction to 3+1 {{Numerical Relativity}}}.
\newblock {Oxford Science Publications}, 2008.

\bibitem{anderson2018a}
M.~T. Anderson.
\newblock On the conformal method for the {{Einstein}} constraint equations.
\newblock 2018.
\newblock Preprint. \href{http://arxiv.org/abs/1812.06320}{arXiv:1812.06320}.

\bibitem{bartnik1986}
R.~A. Bartnik.
\newblock The mass of an asymptotically flat manifold.
\newblock {\em Comm. Pure Appl. Math.}, 39(5):661--693, 1986.
\newblock
  DOI:~\href{https://doi.org/10.1002/cpa.3160390505}{10.1002/cpa.3160390505}.

\bibitem{bartnik1993}
R.~A. Bartnik.
\newblock Quasi-spherical metrics and prescribed scalar curvature.
\newblock {\em J. Diff. Geom.}, 37(1):31--71, 1993.
\newblock
  DOI:~\href{https://doi.org/10.4310/jdg/1214453422}{10.4310/jdg/1214453422}.

\bibitem{bartnik2004}
R.~A. Bartnik and J.~Isenberg.
\newblock The {{Constraint Equations}}.
\newblock In {\em The {{Einstein Equations}} and the {{Large Scale Behavior}}
  of {{Gravitational Fields}}}, pages 1--38. {Birkh{\"a}user Physics}, 2004.

\bibitem{Baumgarte:2010vs}
T.~W. Baumgarte and S.~L. Shapiro.
\newblock {\em Numerical {{Relativity}}}.
\newblock Solving {{Einstein}}'s {{Equations}} on the {{Computer}}. {Cambridge
  University Press}, 2010.

\bibitem{Beyer:2009vw}
F.~Beyer.
\newblock A spectral solver for evolution problems with spatial
  {$S^3$}-topology.
\newblock {\em J. Comp. Phys.}, 228(17):6496--6513, 2009.
\newblock
  DOI:~\href{https://doi.org/10.1016/j.jcp.2009.05.037}{10.1016/j.jcp.2009.05.037}.

\bibitem{Beyer:2015bv}
F.~Beyer, B.~Daszuta, and J.~Frauendiener.
\newblock A spectral method for half-integer spin fields based on spin-weighted
  spherical harmonics.
\newblock {\em Class. Quantum Grav.}, 32(17):175013, 2015.
\newblock
  DOI:~\href{https://doi.org/10.1088/0264-9381/32/17/175013}{10.1088/0264-9381/32/17/175013}.

\bibitem{Beyer:2014bu}
F.~Beyer, B.~Daszuta, J.~Frauendiener, and B.~Whale.
\newblock Numerical evolutions of fields on the 2-sphere using a spectral
  method based on spin-weighted spherical harmonics.
\newblock {\em Class. Quantum Grav.}, 31(7):075019, 2014.
\newblock
  DOI:~\href{https://doi.org/10.1088/0264-9381/31/7/075019}{10.1088/0264-9381/31/7/075019}.

\bibitem{Beyer:2016fc}
F.~Beyer, L.~Escobar, and J.~Frauendiener.
\newblock Numerical solutions of {{Einstein}}'s equations for cosmological
  spacetimes with spatial topology {$S^3$} and symmetry group {$U(1)$}.
\newblock {\em Phys. Rev. D}, 93(4):043009, 2016.
\newblock
  DOI:~\href{https://doi.org/10.1103/PhysRevD.93.043009}{10.1103/PhysRevD.93.043009}.

\bibitem{Beyer:2017tu}
F.~Beyer, L.~Escobar, and J.~Frauendiener.
\newblock Asymptotics of solutions of a hyperbolic formulation of the
  constraint equations.
\newblock {\em Class. Quantum Grav.}, 34(20):205014, 2017.
\newblock
  DOI:~\href{https://doi.org/10.1088/1361-6382/aa8be6}{10.1088/1361-6382/aa8be6}.

\bibitem{Beyer:2017jw}
F.~Beyer, L.~Escobar, and J.~Frauendiener.
\newblock Criticality of inhomogeneous {{Nariai}}-like cosmological models.
\newblock {\em Phys. Rev. D}, 95(8):084030, 2017.
\newblock
  DOI:~\href{https://doi.org/10.1103/PhysRevD.95.084030}{10.1103/PhysRevD.95.084030}.

\bibitem{bieri2009extensions}
L.~Bieri and N.~Zipser.
\newblock {\em Extensions of the Stability Theorem of the {{Minkowski}} Space
  in General Relativity}, volume~45.
\newblock {American Mathematical Society}, 2009.

\bibitem{Bishop:2004gb}
N.~T. Bishop, F.~Beyer, and M.~Koppitz.
\newblock Black hole initial data from a nonconformal decomposition.
\newblock {\em Phys. Rev. D}, 69(6):325, 2004.
\newblock
  DOI:~\href{https://doi.org/10.1103/PhysRevD.69.064010}{10.1103/PhysRevD.69.064010}.

\bibitem{Bishop:1998cb}
N.~T. Bishop, R.~Isaacson, M.~Maharaj, and J.~Winicour.
\newblock Black hole data via a {{Kerr}}-{{Schild}} approach.
\newblock {\em Phys. Rev. D}, 57(10):6113--6118, 1998.
\newblock
  DOI:~\href{https://doi.org/10.1103/PhysRevD.57.6113}{10.1103/PhysRevD.57.6113}.

\bibitem{ChoquetBruhat:1969cl}
Y.~{Choquet-Bruhat} and R.~P. Geroch.
\newblock Global aspects of the {{Cauchy}} problem in general relativity.
\newblock {\em Commun. Math. Phys.}, 14(4):329--335, 1969.
\newblock DOI:~\href{https://doi.org/10.1007/BF01645389}{10.1007/BF01645389}.

\bibitem{klainerman1993global}
D.~Christodoulou and S.~Klainerman.
\newblock {\em The {{Global Nonlinear Stability}} of the {{Minkowski Space}}},
  volume~41 of {\em Princeton {{Mathematical Series}}}.
\newblock {Princeton University Press}, {Princeton}, 1994.

\bibitem{chu2014}
T.~Chu.
\newblock Including realistic tidal deformations in binary black-hole initial
  data.
\newblock {\em Phys. Rev. D}, 89(6):064062, 2014.
\newblock
  DOI:~\href{https://doi.org/10.1103/PhysRevD.89.064062}{10.1103/PhysRevD.89.064062}.

\bibitem{Dain:2002cm}
S.~Dain.
\newblock Black hole interaction energy.
\newblock {\em Phys. Rev. D}, 66(8):084019, 2002.
\newblock
  DOI:~\href{https://doi.org/10.1103/PhysRevD.66.084019}{10.1103/PhysRevD.66.084019}.

\bibitem{Dain:2001cd}
S.~Dain and H.~Friedrich.
\newblock Asymptotically {{Flat Initial Data}} with {{Prescribed Regularity}}
  at {{Infinity}}.
\newblock {\em Commun. Math. Phys.}, 222(3):569--609, 2001.
\newblock
  DOI:~\href{https://doi.org/10.1007/s002200100524}{10.1007/s002200100524}.

\bibitem{dilts2017}
J.~Dilts, M.~Holst, T.~Kozareva, and D.~Maxwell.
\newblock Numerical {{Bifurcation Analysis}} of the {{Conformal Method}}.
\newblock 2017.
\newblock Preprint. \href{http://arxiv.org/abs/1710.03201}{arXiv:1710.03201}.

\bibitem{doulis2019}
G.~Doulis.
\newblock Construction of high precision numerical single and binary black hole
  initial data.
\newblock {\em Phys. Rev. D}, 100(2):024064, 2019.
\newblock
  DOI:~\href{https://doi.org/10.1103/PhysRevD.100.024064}{10.1103/PhysRevD.100.024064}.

\bibitem{FouresBruhat:1952ji}
Y.~{Four{\`e}s-Bruhat}.
\newblock Th{\'e}or{\`e}me d'existence pour certains syst{\`e}mes
  d'{\'e}quations aux d{\'e}riv{\'e}es partielles non lin{\'e}aires.
\newblock {\em Acta Math.}, 88(1):141--225, 1952.
\newblock DOI:~\href{https://doi.org/10.1007/BF02392131}{10.1007/BF02392131}.

\bibitem{Gourgoulhon:Book}
E.~Gourgoulhon.
\newblock {\em 3+1 {{Formalism}} in {{General Relativity}}}, volume 846 of {\em
  Lecture {{Notes}} in {{Physics}}}.
\newblock {Springer Berlin Heidelberg}, {Berlin, Heidelberg}, 2012.

\bibitem{lovelace2009}
G.~Lovelace.
\newblock Reducing spurious gravitational radiation in binary-black-hole
  simulations by using conformally curved initial data.
\newblock {\em Class. Quantum Grav.}, 26(11):114002, 2009.
\newblock
  DOI:~\href{https://doi.org/10.1088/0264-9381/26/11/114002}{10.1088/0264-9381/26/11/114002}.

\bibitem{Matzner:1998hv}
R.~A. Matzner, M.~F. Huq, and D.~Shoemaker.
\newblock Initial data and coordinates for multiple black hole systems.
\newblock {\em Phys. Rev. D}, 59(2):024015, 1998.
\newblock
  DOI:~\href{https://doi.org/10.1103/PhysRevD.59.024015}{10.1103/PhysRevD.59.024015}.

\bibitem{Misner:1973vb}
C.~W. Misner, K.~S. Thorne, and J.~A. Wheeler.
\newblock {\em Gravitation}.
\newblock {Freeman}, 1973.

\bibitem{Moreno:2002dm}
C.~Moreno, D.~N{\'u}{\~n}ez, and O.~Sarbach.
\newblock Kerr\textendash{{Schild}}-type initial data for black holes with
  angular momenta.
\newblock {\em Class. Quantum Grav.}, 19(23):6059--6073, 2002.
\newblock
  DOI:~\href{https://doi.org/10.1088/0264-9381/19/23/312}{10.1088/0264-9381/19/23/312}.

\bibitem{Murchadha:2005ib}
N.~{\'O}. Murchadha and N.~Xie.
\newblock Flat slices in {{Minkowski}} space.
\newblock {\em Class. Quantum Grav.}, 32(6):067001, 2015.
\newblock
  DOI:~\href{https://doi.org/10.1088/0264-9381/32/6/067001}{10.1088/0264-9381/32/6/067001}.

\bibitem{Nakonieczna:2017vk}
A.~Nakonieczna, {\L}.~Nakonieczny, and I.~R{\'a}cz.
\newblock Black hole initial data by numerical integration of the
  parabolic-hyperbolic form of the constraints.
\newblock 2017.
\newblock Preprint. \href{http://arxiv.org/abs/1712.00607}{arXiv:1712.00607}.

\bibitem{Penrose:1984tf}
R.~Penrose and W.~Rindler.
\newblock {\em Two-{{Spinor Calculus}} and {{Relativistic Fields}}}, volume~1
  of {\em Spinors and {{Space}}-{{Time}}}.
\newblock {Cambridge University Press}, {Cambridge}, 1984.

\bibitem{Racz:2014kk}
I.~R{\'a}cz.
\newblock Cauchy problem as a two-surface based `geometrodynamics'.
\newblock {\em Class. Quantum Grav.}, 32(1):015006, 2014.
\newblock
  DOI:~\href{https://doi.org/10.1088/0264-9381/32/1/015006}{10.1088/0264-9381/32/1/015006}.

\bibitem{Racz:2014dx}
I.~R{\'a}cz.
\newblock Is the {{Bianchi}} identity always hyperbolic?
\newblock {\em Class. Quantum Grav.}, 31(15):155004, 2014.
\newblock
  DOI:~\href{https://doi.org/10.1088/0264-9381/31/15/155004}{10.1088/0264-9381/31/15/155004}.

\bibitem{Racz:2015gb}
I.~R{\'a}cz.
\newblock Constraints as evolutionary systems.
\newblock {\em Class. Quantum Grav.}, 33(1):015014, 2016.
\newblock
  DOI:~\href{https://doi.org/10.1088/0264-9381/33/1/015014}{10.1088/0264-9381/33/1/015014}.

\bibitem{racz2018}
I.~R{\'a}cz.
\newblock On the {{Evolutionary Form}} of the {{Constraints}} in
  {{Electrodynamics}}.
\newblock {\em Symmetry}, 11(1):10, 2018.
\newblock DOI:~\href{https://doi.org/10.3390/sym11010010}{10.3390/sym11010010}.

\bibitem{Racz:2015bu}
I.~R{\'a}cz and J.~Winicour.
\newblock Black hole initial data without elliptic equations.
\newblock {\em Phys. Rev. D}, 91(12):124013, 2015.
\newblock
  DOI:~\href{https://doi.org/10.1103/PhysRevD.91.124013}{10.1103/PhysRevD.91.124013}.

\bibitem{winicourRacz2018}
I.~R{\'a}cz and J.~Winicour.
\newblock Toward computing gravitational initial data without elliptic solvers.
\newblock {\em Class. Quantum Grav.}, 35(13):135002, 2018.
\newblock
  DOI:~\href{https://doi.org/10.1088/1361-6382/aac5c5}{10.1088/1361-6382/aac5c5}.

\bibitem{Szabados:2009ig}
L.~B. Szabados.
\newblock Quasi-{{Local Energy}}-{{Momentum}} and {{Angular Momentum}} in
  {{General Relativity}}.
\newblock {\em Living Rev. Relativity}, 12(4):76, 2009.
\newblock DOI:~\href{https://doi.org/10.12942/lrr-2009-4}{10.12942/lrr-2009-4}.

\bibitem{webber2012}
J.~B.~W. Webber.
\newblock A bi-symmetric log transformation for wide-range data.
\newblock {\em Meas. Sci. Technol.}, 24(2):027001, 2012.
\newblock
  DOI:~\href{https://doi.org/10.1088/0957-0233/24/2/027001}{10.1088/0957-0233/24/2/027001}.

\bibitem{winicour2017}
J.~Winicour.
\newblock The algebraic-hyperbolic approach to the linearized gravitational
  constraints on a {{Minkowski}} background.
\newblock {\em Class. Quantum Grav.}, 34(15):157001, 2017.
\newblock
  DOI:~\href{https://doi.org/10.1088/1361-6382/aa7bd6}{10.1088/1361-6382/aa7bd6}.

\end{thebibliography}

\end{document}